\def\complexNumbers{\mathbb{C}}
\def\realNumbers{\mathbb{R}}
\def\figuresize{\textwidth/2-0.3in}

\def\expectationOperator[#1][#2]{{\mathbb{E}_{#2}}\left[#1\right]}
\def\indicatorFunction[#1]{\mathbb{I}\left[{#1}\right]}
\def\probability[#1]{\mathbb{P}\left({#1}\right)}
\def\complexGaussian[#1][#2]{\mathcal{CN}({#1,#2})}
\def\gaussian[#1][#2]{\mathcal{N}({#1,#2})}

\def\normalPDF[#1]{\phi\left(#1\right)}
\def\normalCDF[#1]{\Phi\left(#1\right)}

\def\coeffErr[#1]{A(#1)}
\def\hyperGeometricFcn[#1][#2][#3][#4]{{_2F_1}\left(#1,#2;#3;#4\right)}
\def\numberOfEdgeDevices{K}
\def\timeDomainOFDM[#1]{s(#1)}

\def\numberOfActiveSubcarriers{M}
\def\idftSize{N_{\rm IDFT}}
\def\dataSymbols[#1]{d_{#1}}
\def\cpSize{N_{\rm cp}}

\def\receivedSymbolAtSubcarrier[#1]{r_{#1}^{(\indexCommunicationRound)}}
\def\transmittedSymbolAtSubcarrier[#1]{t_{#1}^{(\indexCommunicationRound)}}
\def\randomSymbolAtSubcarrier[#1]{s_{#1}^{(\indexCommunicationRound)}}
\def\channelAtSubcarrier[#1]{h_{#1}^{(\indexCommunicationRound)}}
\def\noiseAtSubcarrier[#1]{n_{#1}^{(\indexCommunicationRound)}}
\def\numberOfOFDMSymbols{S}
\def\indexOFDMSymbol{m}
\def\asymbolFromED[#1]{d_{#1}}

\def\exponentialIntegral[#1]{{\rm Ei}(#1)}
\def\tciFactor[#1]{p_{#1}}

\def\encoder[#1]{\psi_#1}
\def\symbolEnergy{E_{\rm s}}
\def\voteInTime[#1]{m^{#1}}
\def\voteInFrequency[#1]{l^{#1}}

\def\correctDecision[#1]{p_{#1}}
\def\zeroDecision[#1]{z_{#1}}

\def\incorrectDecision[#1]{q_{#1}}
\def\aparameterForBer[#1]{\epsilon_{#1}}

\def\probabilityIncorrect[#1]{P^{\rm err}_{#1}}

\def\oneVector[#1]{\textbf{\textrm{1}}_{#1}}
\def\zeroVector[#1]{\textbf{\textrm{0}}_{#1}}
\def\identityMatrix[#1]{{\textbf{\textrm{I}}_{#1}}}

\def\dataset[#1]{\mathcal{D}_{#1}}

\def\datasetBatch[#1]{\mathcal{\tilde{D}}_{#1}}
\def\batchSize{n_{\rm b}}

\def\completeData{\mathcal{D}}
\def\numberOfModelParameters{Q}
\def\sampleData[#1]{{\textrm{\textbf{x}}}_{#1}}
\def\sampleLabel[#1]{{y}_{#1}}
\def\learningRate{\eta}

\def\deltaVectorAtIteration[#1][#2]{{\Delta}^{(#1)}_{#2}}
\def\deltaVectorAtIterationEle[#1]{{\bm \Delta}^{#1}}

\def\indexED{k}
\def\indexGradient{q}
\def\indexSampleData{{\ell}}
\def\indexCommunicationRound{n}

\def\modelParametersAtIteration[#1]{\textbf{w}^{(#1)}}
\def\modelParametersAtIterationEle[#1][#2]{w^{(#1)}_{#2}}

\def\modelParameters{\textbf{w}}
\def\modelParametersEle[#1]{{w}_{#1}}
\def\modelParametersOptimal{{\textbf{w}^{*}}}

\def\localGradientSign[#1][#2]{\bar{\textbf{g}}_{#1}^{(#2)}}
\def\localGradient[#1][#2]{\tilde{\textbf{g}}_{#1}^{(#2)}}
\def\localGradientNoIndex[#1]{\tilde{\textbf{g}}_{#1}}

\def\localGradientSignElement[#1][#2]{{\bar{g}}_{#1}^{(#2)}}
\def\localGradientElement[#1][#2]{{\tilde{g}}_{#1}^{(#2)}}
\def\localGradientNoIndexElement[#1]{{\tilde{g}}_{#1}}
\def\gradientWeight[#1]{\omega{\left(#1\right)}}

\def\encoderGradient[#1][#2][#3]{\psi_{#1}^{#2}(#3)}

\def\lossFunctionSample[#1]{f(#1)}
\def\lossFunctionLocal[#1][#2]{F_{#1}(#2)}
\def\lossFunctionGlobal[#1]{F(#1)}

\def\majorityVoteEle[#1][#2]{{v}^{(#1)}_{#2}}
\def\majorityVote[#1]{\textbf{v}^{(#1)}}

\def\majorityVoteSchemeEle[#1][#2]{{\tilde{v}}^{(#1)}_{#2}}
\def\majorityVoteScheme[#1]{\tilde{\textbf{v}}^{(#1)}}

\def\globalGradient[#1]{{\textbf{{g}}}^{(#1)}}
\def\globalGradientElement[#1][#2]{{{g}}^{(#1)}_{#2}}

\def\globalGradientElementNoIndex[#1]{{g_{#1}}}

\def\metricForFirst[#1]{e_{#1}^{+}}
\def\metricForSecond[#1]{e_{#1}^{-}}

\def\nonnegativeConstantsEle[#1]{L_{#1}}

\def\varianceBoundEle[#1]{\sigma_{#1}}

\def\channelVector[#1]{\textbf{\textrm{h}}_{#1}^{(\indexCommunicationRound)}}

\def\receiveVector[#1]{\textbf{\textrm{r}}_{#1}^{(\indexCommunicationRound)}}
\def\receiveVectorEstimate[#1]{\tilde{\textbf{\textrm{{r}}}}_{#1}^{(\indexCommunicationRound)}}

\def\matrixForCut[#1]{\textbf{\textrm{C}}_{#1}}
\def\symbolVector[#1]{\textbf{\textrm{d}}}
\def\symbolVectorEstimate[#1]{\tilde{\textbf{\textrm{{d}}}}_{#1}^{(\indexCommunicationRound)}}
\def\receivedVector[#1]{\textbf{\textrm{r}}_{#1}^{(\indexCommunicationRound)}}
\def\noiseVector[#1]{\textbf{\textrm{n}}_{#1}^{(\indexCommunicationRound)}}
\def\noiseVectorOnSymbols[#1]{\tilde{\textbf{\textrm{n}}}_{#1}^{(\indexCommunicationRound)}}

\def\transmittedVector[#1]{\textbf{\textrm{t}}}
\def\channelVector[#1]{\textbf{\textrm{h}}_{#1}^{(\indexCommunicationRound)}}
\def\cpAddMatrix{\textbf{\textrm{A}}}
\def\idftMatrix[#1]{\textbf{\textrm{F}}_{#1}^{\rm H}}
\def\dftMatrix[#1]{\textbf{\textrm{F}}_{#1}}
\def\transformPrecoder[#1]{\textbf{\textrm{T}}_{#1}}
\def\transformDecoder[#1]{\textbf{\textrm{T}}_{#1}^{\rm H}}
\def\dftPrecoder[#1]{\textbf{\textrm{D}}_{#1}}
\def\dftDecoder[#1]{\textbf{\textrm{D}}_{#1}^{\rm H}}

\def\frequencyMapping{\textbf{\textrm{M}}_{\textrm{f}}}

\def\channelImpulseResponse[#1]{\textbf{\textrm{h}}_{#1}^{(\indexCommunicationRound)}}
\def\channelMatrix[#1]{\textbf{\textrm{H}}_{#1}^{(\indexCommunicationRound)}}
\def\channelMatrixDiag[#1]{{\bf \Lambda}_{#1}^{(\indexCommunicationRound)}}

\def\sampleRate{f_{\rm sample}}
\def\samplePeriod{T_{\rm sample}}

\def\distanceED[#1]{r_{#1}}
\def\powerED[#1]{P_{#1}}

\def\thresholdForZero{\mathscr{t}}

\newcommand\mydots{\hbox to 1em{.\hss.\hss.}}

\documentclass[conference,comsoc]{IEEEtran}

\usepackage{multicol}
\usepackage{lipsum}
\usepackage{mathtools}
\usepackage{amsthm}
\usepackage{acronym}
\usepackage[bookmarksopen=true]{hyperref}
\usepackage{stfloats}
\usepackage{amsfonts}
\usepackage{cite}
\usepackage{multirow}
\usepackage{bm}
\usepackage{amsmath,amssymb}
\usepackage{graphicx}
\usepackage[table,xcdraw]{xcolor}
\usepackage[caption=false,font=footnotesize]{subfig}
\usepackage[geometry]{ifsym}
\usepackage{array}
\usepackage[utf8]{inputenc}
\usepackage[T1]{fontenc}

\let\norm\undefined 
\DeclarePairedDelimiter\norm{\lVert}{\rVert}
\DeclarePairedDelimiter\inprod{\langle}{\rangle}

\usepackage{tikz}
\usepackage{lipsum}
\usetikzlibrary{calc,shadings,patterns}
\makeatletter
\tikzset{%
  remember picture with id/.style={%
    remember picture,
    overlay,
    save picture id=#1,
  },
  save picture id/.code={%
    \edef\pgf@temp{#1}%
    \immediate\write\pgfutil@auxout{%
      \noexpand\savepointas{\pgf@temp}{\pgfpictureid}}%
  },
  if picture id/.code args={#1#2#3}{%
    \@ifundefined{save@pt@#1}{%
      \pgfkeysalso{#3}%
    }{
      \pgfkeysalso{#2}%
    }
  }
}

\def\savepointas#1#2{%
  \expandafter\gdef\csname save@pt@#1\endcsname{#2}%
}

\def\tmk@labeldef#1,#2\@nil{%
  \def\tmk@label{#1}%
  \def\tmk@def{#2}%
}

\tikzdeclarecoordinatesystem{pic}{%
  \pgfutil@in@,{#1}%
  \ifpgfutil@in@%
    \tmk@labeldef#1\@nil
  \else
    \tmk@labeldef#1,(0pt,0pt)\@nil
  \fi
  \@ifundefined{save@pt@\tmk@label}{%
    \tikz@scan@one@point\pgfutil@firstofone\tmk@def
  }{%
  \pgfsys@getposition{\csname save@pt@\tmk@label\endcsname}\save@orig@pic%
  \pgfsys@getposition{\pgfpictureid}\save@this@pic%
  \pgf@process{\pgfpointorigin\save@this@pic}%
  \pgf@xa=\pgf@x
  \pgf@ya=\pgf@y
  \pgf@process{\pgfpointorigin\save@orig@pic}%
  \advance\pgf@x by -\pgf@xa
  \advance\pgf@y by -\pgf@ya
  }%
}

\makeatother

\newcounter{hatchNumber}
\DeclarePairedDelimiter\ceil{\lceil}{\rceil}

\usepackage{cuted}
\makeatletter
\newif\ifAC@uppercase@first%
\def\Aclp#1{\AC@uppercase@firsttrue\aclp{#1}\AC@uppercase@firstfalse}%
\def\AC@aclp#1{%
	\ifcsname fn@#1@PL\endcsname%
	\ifAC@uppercase@first%
	\expandafter\expandafter\expandafter\MakeUppercase\csname fn@#1@PL\endcsname%
	\else%
	\csname fn@#1@PL\endcsname%
	\fi%
	\else%
	\AC@acl{#1}s%
	\fi%
}%
\def\Acp#1{\AC@uppercase@firsttrue\acp{#1}\AC@uppercase@firstfalse}%
\def\AC@acp#1{%
	\ifcsname fn@#1@PL\endcsname%
	\ifAC@uppercase@first%
	\expandafter\expandafter\expandafter\MakeUppercase\csname fn@#1@PL\endcsname%
	\else%
	\csname fn@#1@PL\endcsname%
	\fi%
	\else%
	\AC@ac{#1}s%
	\fi%
}%
\def\Acfp#1{\AC@uppercase@firsttrue\acfp{#1}\AC@uppercase@firstfalse}%
\def\AC@acfp#1{%
	\ifcsname fn@#1@PL\endcsname%
	\ifAC@uppercase@first%
	\expandafter\expandafter\expandafter\MakeUppercase\csname fn@#1@PL\endcsname%
	\else%
	\csname fn@#1@PL\endcsname%
	\fi%
	\else%
	\AC@acf{#1}s%
	\fi%
}%
\def\Acsp#1{\AC@uppercase@firsttrue\acsp{#1}\AC@uppercase@firstfalse}%
\def\AC@acsp#1{%
	\ifcsname fn@#1@PL\endcsname%
	\ifAC@uppercase@first%
	\expandafter\expandafter\expandafter\MakeUppercase\csname fn@#1@PL\endcsname%
	\else%
	\csname fn@#1@PL\endcsname%
	\fi%
	\else%
	\AC@acs{#1}s%
	\fi%
}%
\edef\AC@uppercase@write{\string\ifAC@uppercase@first\string\expandafter\string\MakeUppercase\string\fi\space}%
\def\AC@acrodef#1[#2]#3{%
	\@bsphack%
	\protected@write\@auxout{}{%
		\string\newacro{#1}[#2]{\AC@uppercase@write #3}%
	}\@esphack%
}%
\def\Acl#1{\AC@uppercase@firsttrue\acl{#1}\AC@uppercase@firstfalse}
\def\Acf#1{\AC@uppercase@firsttrue\acf{#1}\AC@uppercase@firstfalse}
\def\Ac#1{\AC@uppercase@firsttrue\ac{#1}\AC@uppercase@firstfalse}
\def\Acs#1{\AC@uppercase@firsttrue\acs{#1}\AC@uppercase@firstfalse}

\DeclareMathOperator{\sign}{sign}
\DeclareMathOperator{\diag}{diag}
\def\diagOperator[#1]{\diag\left(#1\right)}
\def\signNormal[#1]{\sign\left(#1\right)}
\def\signThreshold[#1][#2]{\sign_{#2}\left(#1\right)}

\acrodef{WSN}{wireless sensor network}
\acrodef{SN}{sensor node}
\acrodef{FC}{fusion center}
\acrodef{MAC}{multiple-access channel}
\acrodef{FL}{federated learning}
\acrodef{ED}{edge device}
\acrodef{CS}{compressed sensing}
\acrodef{ES}{edge server}
\acrodef{AXI}{advanced extensible interface}
\acrodef{CC}{companion computer}
\acrodef{CFO}{carrier frequency offset}
\acrodef{RRC}{root-raised cosine}
\acrodef{CHEST}{channel estimation}
\acrodef{PPDU}{physical layer protocol data unit}
\acrodef{DC}{direct current}
\acrodef{CRC}{cyclic redundancy check}

\acrodef{ZC}{Zadoff-Chu}

\acrodef{STLC}{space-time line code}
\acrodef{CCI}{co-channel interference}
\acrodef{CSIT}[CSIT]{\ac{CSI} at the transmitter}
\acrodef{CSIR}[CSIR]{\ac{CSI} at the receiver}
\acrodef{MIMO}{multiple-input-multiple-output}
\acrodef{PC}{phase correction}
\acrodef{ZF}{zero-forcing}
\acrodef{ANOVA}{analysis of variance}
\acrodef{SDR}{software-defined radio}
\acrodef{PS}{processing system}
\acrodef{SS}{soft synchronization}
\acrodef{IQ}{in-phase/quadrature}
\acrodef{IP}{intellectual property}
\acrodef{DMA}{direct-memory access}
\acrodef{RAM}{random access memory}
\acrodef{UE}{user equipment}
\acrodef{STA}{station}
\acrodef{BS}{base station}
\acrodef{MS}{mobile station}
\acrodef{MSE}{mean squared error}
\acrodef{TDMA}{time-domain multiple access}
\acrodef{GPS}{Global Positioning System}

\acrodef{FPGA}{field-programmable gate array}

\acrodef{FSK}{frequency-shift keying}
\acrodef{PPM}{pulse-position modulation}
\acrodef{PAM}{pulse-amplitude modulation}

\acrodef{MRC}{maximum-ratio combining}
\acrodef{HP}{hard-coded participation}
\acrodef{HPA}{hard-coded participation with absentees}
\acrodef{SP}{soft-coded participation}
\acrodef{FSK-MV}{\ac{FSK}-based \ac{MV}}
\acrodef{RF}{radio-frequency}
\acrodef{MF}{matched filter}
\acrodef{PPM}{pulse-position modulation}
\acrodef{CSK}{chirp-shift keying}
\acrodef{PPM-MV}[PPM-MV]{\ac{PPM}-based \ac{MV}}
\acrodef{DFT-s-OFDM}{\ac{DFT}-spread \ac{OFDM}}
\acrodef{SC}{single-carrier}
\acrodef{SGD}{stochastic gradient descent}
\acrodef{signSGD}{sign stochastic gradient descent}

\acrodef{SL}{split learning}
\acrodef{SNR}{signal-to-noise ratio}
\acrodef{RMSE}{root-mean-square error}
\acrodef{OFDM}{orthogonal frequency division multiplexing}
\acrodef{DFT}{discrete Fourier transform}
\acrodef{PSK}{phase-shift keying}
\acrodef{QAM}{quadrature amplitude modulation}
\acrodef{QPSK}{quadrature phase-shift keying}
\acrodef{PMEPR}{peak-to-mean envelope power ratio}
\acrodef{BER}{bit-error ratio}
\acrodef{SNR}{signal-to-noise ratio}
\acrodef{PSD}{power spectral density}
\acrodef{SE}{spectral efficiency}
\acrodef{CP}{cyclic prefix}
\acrodef{AWGN}{additive white Gaussian noise}
\acrodef{CFR}{channel frequency response}
\acrodef{CIR}{channel impulse response}
\acrodef{MMSE}{minimum mean square error}
\acrodef{LMMSE}{linear minimum mean square error}
\acrodef{BPSK}{binary phase shift keying}
\acrodef{BLER}{block-error rate}
\acrodef{ML}{maximum likelihood}
\acrodef{PHY}{physical layer}
\acrodef{PA}{power amplifier}
\acrodef{IDFT}{inverse DFT}
\acrodef{DoF}{degrees-of-freedom}
\acrodef{IoT}{Internet-of-Things}
\acrodef{FDE}{frequency-domain equalization}
\acrodef{RF}{radio-frequency}
\acrodef{IM}{index modulation}
\acrodef{BS}{base station}
\acrodef{MF}{matched filter}
\acrodef{PPM}{pulse-position modulation}

\acrodef{MSE}{mean-square error}
\acrodef{MRT}{maximum-ratio transmission}
\acrodef{ERC}{equal-ratio combining}
\acrodef{BAA}{broadband analog aggregation}
\acrodef{OBDA}{one-bit broadband digital aggregation}
\acrodef{FEEL}{federated edge learning}
\acrodef{FL}{federated learning}
\acrodef{ED}{edge device}
\acrodef{ES}{edge server}
\acrodef{UL}{uplink}
\acrodef{DL}{downlink}
\acrodef{OAC}{over-the-air computation}
\acrodef{TCI}{truncated-channel inversion}
\acrodef{MV}{majority vote}
\acrodef{CNN}{convolution neural network}
\acrodef{ReLU}{rectified-linear unit}
\acrodef{CSI}{channel state information}
\acrodef{PAPR}{peak-to-average power ratio}
\acrodef{SC}{single-carrier}
\acrodef{iid}[IID]{independent and identically distributed}
\acrodef{RMS}{root-mean-square}
\acrodef{4G}{Fourth Generation}
\acrodef{5G}{Fifth Generation}
\acrodef{NR}{New Radio}
\acrodef{LTE}{Long-Term Evolution}
\acrodef{DFT-s-OFDM}{\ac{DFT}-spread \ac{OFDM}}
\acrodef{OFDMA}{orthogonal frequency division multiple access}
\acrodef{HARQ}{hybrid automatic repeat request}
\acrodef{D2D}{Device-to-Device}
\acrodef{NOMA}{non-orthogonal multiple access}
\def\BibTeX{{\rm B\kern-.05em{\sc i\kern-.025em b}\kern-.08em
    T\kern-.1667em\lower.7ex\hbox{E}\kern-.125emX}}
\begin{document}

\title{A Demonstration of Over-the-Air Computation for Federated Edge Learning} 

\author{
	\IEEEauthorblockN{Alphan \c{S}ahin} \IEEEauthorblockA{Electrical Engineering Department, University of South Carolina, Columbia, SC, USA\\
		Email: asahin@mailbox.sc.edu}
} 

\maketitle

\begin{abstract}
In this study, we propose a general-purpose synchronization method that allows a set of \acp{SDR} to transmit or receive any \acl{IQ} data with precise timings while maintaining the baseband processing in the corresponding \aclp{CC}. The proposed method relies on the detection of a synchronization waveform in both receive and transmit directions and  controlling the \acl{DMA} blocks jointly with the processing system.  
By implementing this synchronization method on a set of low-cost \acp{SDR}, we demonstrate the performance of \acl{FSK-MV}, i.e., an \acl{OAC} scheme for \acl{FEEL},   and introduce the corresponding procedures. Our experiment shows that the test accuracy can reach more than 95\%  for homogeneous and heterogeneous data distributions without using channel state information at the \aclp{ED}.
\end{abstract}
\section{Introduction}

\acresetall



\Ac{OAC} leverages the signal-superposition property of wireless multiple-access channels to compute a nomographic function \cite{goldenbaum2015nomographic}. It has recently gained major attention to reduce the per-round communication latency that linearly increases with the number of \acp{ED}  for \ac{FEEL}, i.e., an implementation of \ac{FL} in a wireless network \cite{Guangxu_2021,liu2021training}. Despite its merit,  an \ac{OAC} scheme may require the \acp{ED} to start their transmissions synchronously with high accuracy \cite{Altun_2021survey}, which can impose stringent requirements for the underlying mechanisms. In a practical network, time synchronization can be maintained via an external timing reference such as the \ac{GPS} (see \cite{Alemdar_2021} and the references therein), a triggering mechanism   as in IEEE 802.11 \cite{BELLALTA2019145}, or  well-designed synchronization procedures over random-access and control channels as in cellular networks \cite{10.5555/3294673}. However, while using a \ac{GPS}-based solution can be costly and not suitable for indoor applications, the implementations of a trigger-based synchronization or  some synchronization protocols may not be self-sufficient. This is because an entire baseband  besides the synchronization blocks may need to be implemented as a hard-coded block to satisfy the timing constraints. On the other hand, when a \ac{SDR} is used as  an I/O peripheral  connected to a \ac{CC} for flexible baseband processing, the transmission/reception instants are subject to a large jitter  due to the underlying protocols (e.g., USB, TCP/IP) for the communication between the \ac{CC} and the \ac{SDR}. Hence, it is not trivial to use \acp{SDR} to test an \ac{OAC} scheme in practice.


In the state-of-the-art,  proof-of-concept \ac{OAC} demonstrations are particularly in the area of wireless sensor networks. For example, in \cite{Jakimovski_2011},  a statistical \ac{OAC} is implemented with twenty-one RFID tags to compute the percentages of the activated classes that encode various temperature ranges. A trigger signal  is used to achieve time synchronization across the RFIDs. In \cite{Kortke_2014}, Goldenbaum and Sta\'nczak's scheme \cite{Goldenbaum_2013tcom} is implemented  with three \acp{SDR} emulating eleven sensor nodes and a fusion center. The arithmetic and geometric mean of the sensor readings are computed over a 5~MHz signal. The time synchronization across the sensor nodes is maintained based on a trigger signal and the proposed method is implemented in a \ac{FPGA}. A calibration procedure is also discussed  to ensure amplitude alignment at the fusion center. In \cite{Alton2017testbed}, the summation is evaluated with a testbed that involves three \acp{SDR} as transmitters and an \ac{SDR} as a receiver. The  scheme used in this setup is based on channel inversion. 
However, the details related to the synchronization are not provided. To the best of our knowledge, an \ac{OAC} scheme has not been demonstrated in practice for \ac{FEEL}. In this study, we address this gap  and introduce a synchronization method  suitable for \acp{SDR}. Our contributions are as follows:


{\bf Synchronization for CC-based baseband processing:}   
To maintain the time synchronization in an \ac{SDR}-based network while maintaining the baseband in the \acp{CC}, we propose a hard-coded block that is agnostic to the \ac{IQ} data desired to be communicated  in the \ac{CC}. We discuss the corresponding procedures, calibration, and synchronization waveform to address the hardware limitations.

{\bf Realization of OAC in practice for FEEL:} We realize the proposed method with an \ac{IP} core embedded into Adalm Pluto \ac{SDR}. 
By using the proposed synchronization method, we demonstrate the performance of \ac{FSK-MV} \cite{sahinCommnet_2021,mohammadICC_2022,sahin2022mv}, i.e., an \ac{OAC} scheme for \ac{FEEL}, for both homogeneous and heterogeneous data distribution scenarios. We also provide the corresponding procedures.

{\em Notation:} The complex  and real numbers are denoted by $\complexNumbers$ and  $\realNumbers$, respectively. 


\def\indexSample{n}
\def\ssSync{\textbf{x}_{\rm SYNC}}
\def\signalTXDL{\textbf{x}_{\rm DL}}
\def\signalTXUL[#1]{\textbf{x}_{{\rm UL},#1}}
\def\iqTX{x_{\rm tx}[\indexSample]}
\def\iqRX{x_{\rm rx}[\indexSample]}
\def\iqTXinput[#1]{x_{\rm tx}[#1]}
\def\iqRXinput[#1]{x_{\rm rx}[#1]}

\def\enTX{e_{\rm tx}[\indexSample]}
\def\enRX{e_{\rm rx}[\indexSample]}
\def\syncDet{e_{\rm det}[\indexSample]}
\def\timeWait[#1]{T_{\rm wait#1}}
\def\timePC[#1]{T_{\rm PC#1}}
\def\timeRX[#1]{T_{\rm RX#1}}
\def\timeTX[#1]{T_{\rm TX#1}}
\def\timeDelta{T_{\Delta}}
\def\NsamplesToBeReceived[#1]{N_{\rm #1}}
\def\Ndl{N_{\rm DL}}
\def\Nul{N_{\rm UL}}
\def\cmdRefill[#1]{\mathop{\mathrm{refill}}({#1})}
\def\cmdRead[#1]{\mathop{\mathrm{read}}({#1})}
\def\cmdTransmit[#1]{\mathop{\mathrm{transmit}}({#1})}
\section{Proposed Synchronization Method}
\begin{figure}
	\centering
	\subfloat[Scenario.]{\includegraphics[width =1.1in]{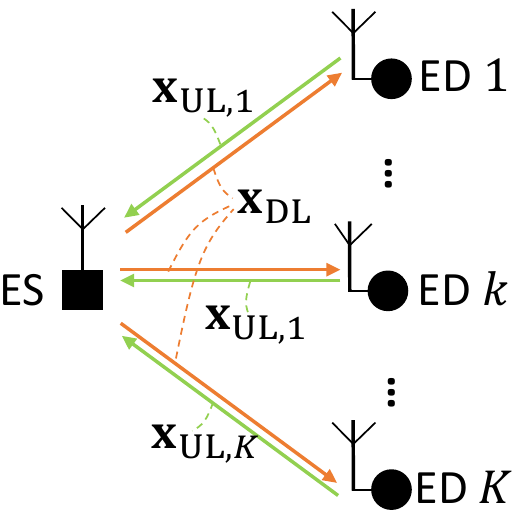}
		\label{subfig:systemModel}}~
	\subfloat[SDR with  the proposed synchronization IP.]{\includegraphics[width =2.1in]{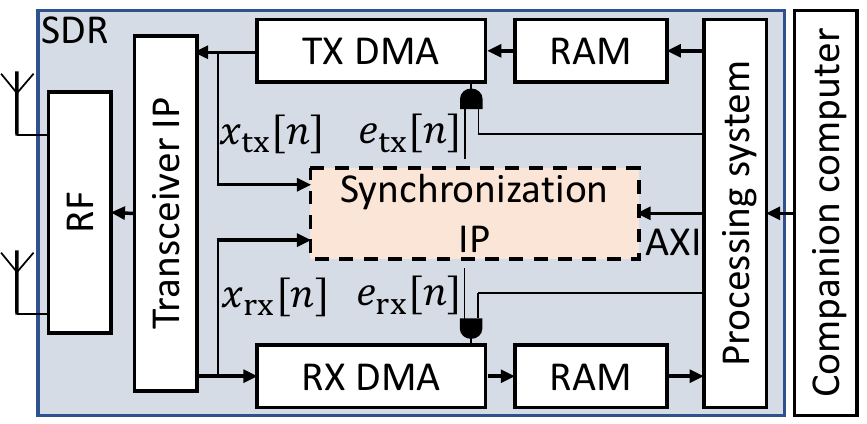}
		\label{subfig:diagram}}
	\\
	\subfloat[The proposed procedure. While there is a large jitter for any transactions between the RAM and the CC, the proposed block ensures precise timings for the reception of $\signalTXDL$ at the EDs, the synchronous transmissions of ${\signalTXUL[1]}$ and ${\signalTXUL[2]}$ to the ES, and the reception of the superposed signal.]{\includegraphics[width = 3.5in]{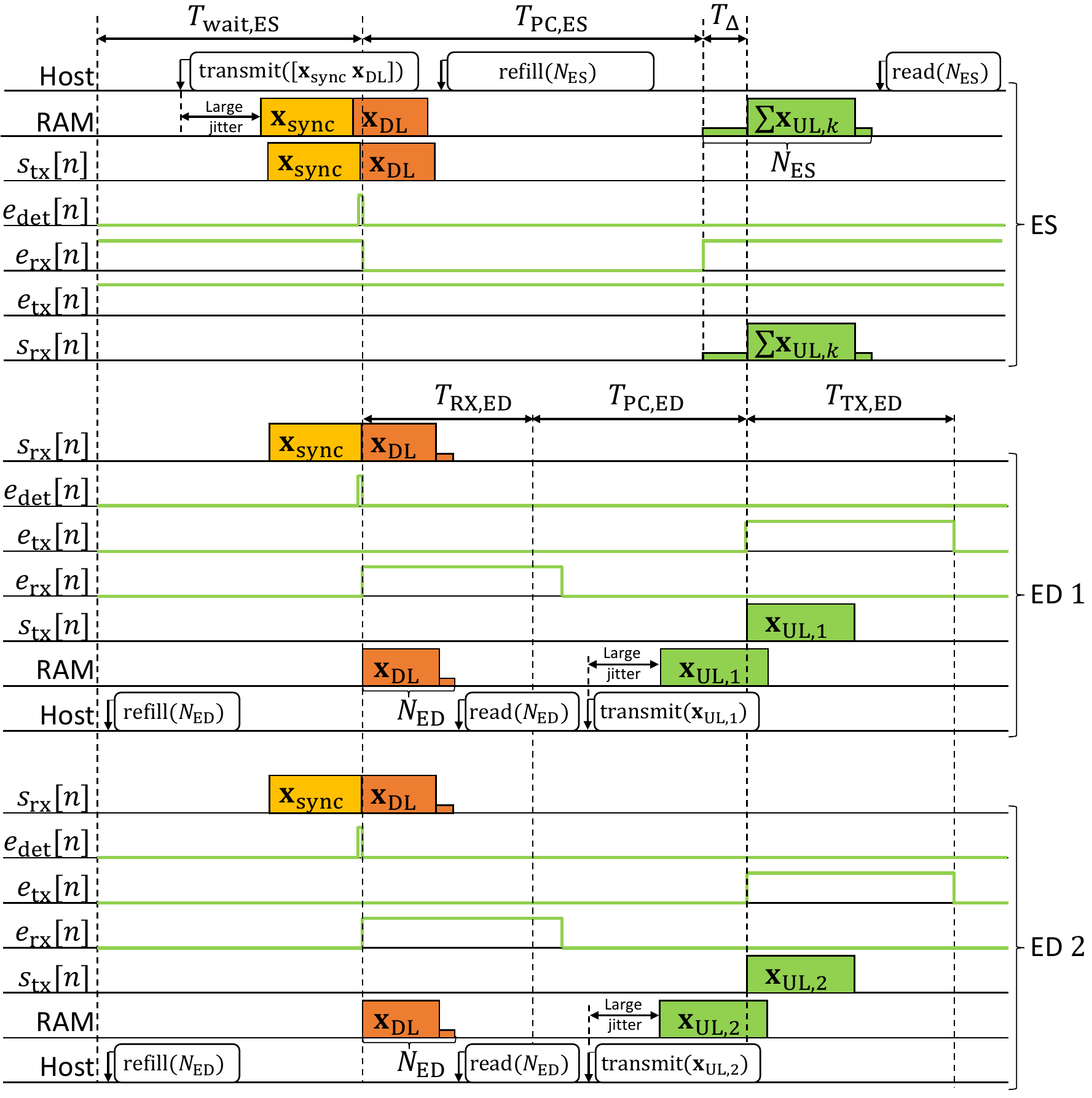}
		\label{subfig:timing}}	
	\caption{The proposed synchronization block and the corresponding procedure.}
	\label{fig:ssSDR}	
\end{figure}

Consider a scenario where $\numberOfEdgeDevices$ \acp{ED} transmit a set of complex-valued vectors denoted by $\{\signalTXUL[\indexED]\in\complexNumbers^{1\times\Nul}|\indexED={1,\mydots,\numberOfEdgeDevices}\}$ to an \ac{ES} in the \ac{UL} in response to the vector $\signalTXDL\in\complexNumbers^{1\times\Ndl}$ transmitted in the \ac{DL} from the \ac{ES}, as illustrated in \figurename~\ref{fig:ssSDR}\subref{subfig:systemModel}. Assume that the implementation of each \ac{ED} (and the \ac{ES}) is based on an \ac{SDR} where the baseband processing is handled by a \ac{CC}. Also,  due to the communication protocol between the \ac{CC} and the \ac{SDR}, consider a large jitter (e.g., in the range of 100 ms) when the IQ data is transferred between  the \ac{CC} and the \ac{SDR}.  Given the large jitter, our goal is to ensure 1) the reception of the vector $\signalTXDL$ at the \ac{CC} of each \ac{ED} and 2) the reception of the superposed vector $\sum_{\indexED=1}\signalTXUL[\indexED]$ (i.e., implying synchronous transmissions
 for simultaneous reception) at the \ac{ES}  with precise timings (e.g., in the order of $\mu s$) while maintaining the baseband at the \acp{CC}. 

To address the scenario above, the main strategy that we adopt  is to separate any signal processing blocks that maintain the synchronization from the ones that do not need to be implemented under strict timing requirements so that the baseband can still be kept in the \ac{CC}. Based on this strategy, we propose a hard-coded block that is solely responsible for time synchronization.
 As shown in \ref{fig:ssSDR}\subref{subfig:diagram}, the proposed block jointly controls the TX  \ac{DMA} and the  RX \ac{DMA}\footnote{TX  \ac{DMA} and RX \ac{DMA} are responsible for transferring  the \ac{IQ} samples from the \ac{RAM} to the transceiver \ac{IP} or vice versa, respectively. They are programmed by the PS, not by the block.} with the \ac{PS} (e.g.,  Linux on the SDR) as a function of the detection of the synchronization waveform, denoted by  $\ssSync$, in the transmit or receive directions through the (active-high) digital signals $\enTX\in\{0,1\}$ and $\enRX\in\{0,1\}$, respectively. 
We define two modes for the block:

\textbf{Mode 1:} The default values of $\enTX$ and $\enRX$ are $0$, i.e., TX \ac{DMA} and RX DMA cannot transfer the \ac{IQ} samples. The block listens to  the output of the transceiver \ac{IP} (i.e., the  \ac{IQ}  samples in the receive direction), denoted by $\iqRX$, and constantly searches for the synchronization waveform  $\ssSync$. If the vector $\ssSync$ is detected, it sequentially sets $(\enTX,\enRX)=(0,1)$ for $\timeRX[]$~seconds to allow the RX DMA to move the received IQ samples to the \ac{RAM}, sets $(\enTX,\enRX)=(0,0)$ for $\timePC[]$~seconds, and finally sets $(\enTX,\enRX)=(1,0)$ for $\timeTX[]$~seconds to allow TX DMA to transfer the IQ  samples from the \ac{RAM} to the transceiver \ac{IP}.

\textbf{Mode 2:} The default values of $\enTX$ and $\enRX$ are $1$, i.e., TX \ac{DMA} and RX DMA can transfer the \ac{IQ} samples. However, the block listens to the output of the TX \ac{DMA} (the  \ac{IQ}  samples  in the transmit direction), denoted by $\iqTX$. It searches for the vector $\ssSync$. If the vector $\ssSync$ is detected, it blocks the reception by setting $\enRX=0$ for $\timePC[]$~seconds. 

Now, assume that the \acp{SDR}  at the \acp{ED} and the \ac{ES} are equipped with the proposed block and operate at Mode 1 and Mode 2, respectively. We propose the following procedure, illustrated in \figurename~\ref{fig:ssSDR}\subref{subfig:timing}, for synchronous communication:

\textbf{Step 1 (EDs)}: The \ac{CC} at each \ac{ED} executes a command (i.e., $\cmdRefill[{\NsamplesToBeReceived[ED]}]$) to fill the \ac{RAM} with  $\NsamplesToBeReceived[ED]$ \ac{IQ} samples  in the receive direction for $\NsamplesToBeReceived[ED]\ge\Ndl$. Since RX \ac{DMA} is disabled by the proposed block at this point, the \ac{CC} waits for the RX \ac{DMA} to be enabled by the block.

\textbf{Step 2 (ES)}:  After the \ac{CC} at the \ac{ES} synthesizes the vector $\signalTXDL$, it prepends $\ssSync$ to initiate the procedure. It writes  $[\ssSync~\signalTXDL]$ to the \ac{RAM} and starts TX DMA  by executing a command (i.e., $\cmdTransmit[{[\ssSync~\signalTXDL]}]$). As soon as the block detects the vector $\ssSync$ in the transmit direction, it disables RX-DMA for $\timePC[,ES]$~seconds. Subsequently, the \ac{CC} issues another command, i.e., $\cmdRefill[{\NsamplesToBeReceived[ES]}]$, to fill its RAM in the receive direction, where $\NsamplesToBeReceived[ES]$ is the number of \ac{IQ} samples to be acquired. However, the reception does not start for $\timePC[,ES]$~seconds due to the disabled RX DMA.
	
\textbf{Step 3 (EDs):}  The transceiver IP at each \ac{ED} receives $[\ssSync~\signalTXDL]$. As soon as the block detects  $\ssSync$, it enables RX  \ac{DMA}. Assuming that $\timeRX[,ED]$ is large enough to acquire $\NsamplesToBeReceived[ED]$ samples, the RX DMA transfers $\NsamplesToBeReceived[ED]$ samples to the RAM  as the \ac{PS} requests for $\NsamplesToBeReceived[ED]$ IQ samples on Step~1. The \ac{CC} reads $\NsamplesToBeReceived[ED]$ IQ samples in the \ac{RAM} via a command, i.e.,  $\cmdRead[{\NsamplesToBeReceived[ED]}]$.
	As a result, $\signalTXDL$ is received with a precise timing.
	
\textbf{Step 4 (EDs)}: The \ac{CC} at the $\indexED$th \ac{ED} processes the vector $\signalTXDL$ and synthesizes $\signalTXUL[\indexED]$ as a response. It then writes $\signalTXUL[\indexED]$ to the RAM and initiates TX \ac{DMA} by executing $\cmdTransmit[{[\ssSync~\signalTXDL]}]$  before the block enables the TX DMA to transfer. Hence,  $\signalTXUL[\indexED]$ should be ready in the RAM within $\timeRX[,ED]+\timePC[,ED]$~seconds. 
	
\textbf{Step 5 (EDs):} The proposed block at the \ac{ED} enables the TX-DMA for $\timeTX[,ED]$~seconds, where $\timeTX[,ED]$ is assumed to be large enough to transmit $\Nul$ IQ samples. At this point, the \acp{ED} start their transmissions simultaneously.
	
\textbf{Step 6 (ES):} Assuming that $\timePC[,ES]=\timeRX[,ED]+\timePC[,ED]-\timeDelta$ and $\NsamplesToBeReceived[ES]\ge\Nul+\ceil{\timeDelta/\samplePeriod}$, the RX DMA at the ES starts to transfer $\NsamplesToBeReceived[ES]$ IQ samples (due to the request in Step 2) $\timeDelta$~second before the \acp{ED}' transmissions, where $\samplePeriod$ is the sample period. After executing  $\cmdRead[{\NsamplesToBeReceived[ES]}]$, the ES receives the superposed signal starting from the $\ceil{\timeDelta/\samplePeriod}$ sample.

The procedure can be repeated after the \ac{ES} waits for $\timeWait[,ES]$~seconds to allow the \acp{ED} to be ready for the next communication cycle and complete its own internal signal processing, where each cycle takes $\timePC[,ED]+\timeRX[,ED]+\timeTX[,ED]+\timeWait[,ES]$~seconds.
 Note that the parameters $\timePC[,ED]$, $\timeRX[,ED]$, $\timeTX[,ED]$, $\timePC[,ES]$, $\timeDelta$ and $\timeWait[,ES]$ can be pre-configured or configured online by the \ac{CC} (e.g., through an \ac{AXI}). Their values depend on the (slowest) processing speed of the constituent \acp{CC} in the network. The timers for $\timePC[,ED]$, $\timeRX[,ED]$, $\timeTX[,ED]$, and $\timePC[,ES]$ can be implemented as counters that count up on each \ac{FPGA} clock tick. The distinct feature of the proposed block and the corresponding procedure is that the timers are set up via $\ssSync$ in the receive and transmit directions at both \acp{ED} and \ac{ES} without using the \ac{CC}.

\subsection{Synchronization Waveform Design and Its Detection}
\label{subsec:syncWaveform}
\def\rollOff{\beta}
\def\Nup{N_{\rm up}}
\def\rrc[#1]{h_{\rm RRC}[#1]}
\def\golaySequnce{\textbf{g}}
\def\golaySequnceEle[#1]{g_{#1}}
\def\xcorr[#1]{\rho[#1]}
\def\acorr[#1]{r[#1]}
\def\metric[#1]{m[#1]}
\def\signalOne[#1]{\textbf{s}_{#1}}
\def\signalTwo{\textbf{b}}

The design of the synchronization waveform  $\ssSync$ and its detection under \ac{CFO} with limited \ac{FPGA} resources were two major issues that we dealt with in our implementation. We address these challenges by synthesizing $\ssSync$ based on a \ac{SC} waveform with the roll-off factor of 0.5  by upsampling a repeated \ac{BPSK} modulated sequence, i.e., $2[\golaySequnce~\golaySequnce~\golaySequnce~\golaySequnce]-1$, by a factor of $\Nup=2$ and passing it through a \ac{RRC} filter,
where $\golaySequnce=[\golaySequnceEle[1],\mydots,\golaySequnceEle[32]]\in\realNumbers^{1\times32}$ is a binary Golay sequence. As a result, the null-to-null bandwidth of  $\ssSync$ is equal to $0.75\sampleRate$, where $\sampleRate$ is the sample rate. 

The rationale behind the design of $\ssSync$ is as follows: 1) An \ac{SC} waveform with a low-order modulation has a small dynamic range. Hence, it requires less power back-off while it can be represented better after the quantization. 
2) A cross-correlation operation can take a large number of \ac{FPGA} resources due to the multiplications. However, the resulting waveform  with the \ac{SC} waveform with a large roll-off factor is similar to the \ac{SC} waveform with a rectangular filter. Hence, we can approximately calculate the normalized cross-correlation by using its approximate \ac{SC} waveform where its samples are either $1$ or $-1$. Hence, the multiplications needed for the cross-correlation can be implemented  with  additions or subtractions. 3) 
In practice, $\ssSync$ is distorted due to the \ac{CFO}. Hence, using a long sequence for cross-correlation can  deteriorate the detection performance. To address this issue, we detect the presence of a shorter sequence, i.e., $\golaySequnce$, back to back four times to declare a detection (i.e., $\syncDet=1$). We choose {\em four} repetitions as it provides a good trade-off between overhead and the detection performance. The metric that we use for the detection of $\golaySequnce$ can be expressed as
\begin{align}
	\metric[\indexSample]\triangleq\frac{1}{\norm{\signalTwo}^2}\frac{|\xcorr[\indexSample]|^2}{|\acorr[\indexSample]|^2}=\frac{1}{\norm{\signalTwo}^2}\frac{\inprod{\signalOne[\indexSample],\signalTwo}^2}{\inprod{\signalOne[\indexSample],\signalOne[\indexSample]}^2} = \frac{\inprod{\signalOne[\indexSample],\signalTwo}^2/2^{12}}{\norm{\signalOne[\indexSample]}^2}
\end{align}
where $\signalTwo$ is based on the approximate \ac{SC} waveform with the rectangular filter and equal to $\signalTwo=2[\golaySequnceEle[32],\golaySequnceEle[32],\golaySequnceEle[31],\golaySequnceEle[31],\mydots,\golaySequnceEle[1],\golaySequnceEle[1]]-1\in\realNumbers^{1\times64}$ for $\Nup=2$ and $\signalOne[\indexSample]$ is
$[\iqRXinput[\indexSample-63],\iqRXinput[\indexSample-62],\mydots,\iqRXinput[\indexSample]]$ for Mode 1 or $[\iqTXinput[\indexSample-63],\iqTXinput[\indexSample-62],\mydots,\iqTXinput[\indexSample]]$ for Mode 2. The block declares a detection if $\metric[\indexSample]$ is larger than 1/4 for four times with 64 samples apart. 

\subsection{Addressing Inaccurate Clocks with Calibration Procedure}
\label{subsec:nonIdealClocks}
\def\fpgaClockPeriod{T_{\rm clk}}
\def\fpgaClockPeriodActual{T^{'}_{{\rm clk,}\indexED}[\indexSample]}
\def\fpgaClockPeriodOffset{\Delta T_{{\rm clk},\indexED}}
\def\fpgaClockPeriodNoise{n_{{\rm clk},\indexED}[\indexSample]}
\def\errorTime[#1]{\Delta T_{#1}}
\def\Ncount[#1]{N_{{\rm cnt}}}
\def\cmdPPDU[#1]{\textbf{t}_{\rm #1}}
\def\signalCal[#1]{\textbf{x}_{{\rm cal},#1}}

\begin{figure}
	\centering
	\subfloat[Calibration trigger.]{\includegraphics[width =3.1in]{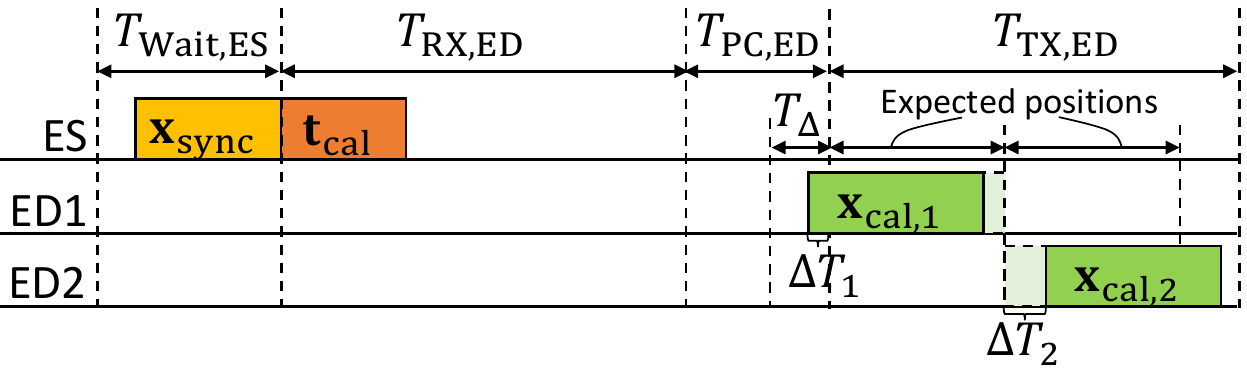}
		\label{subfig:calTrigger}}\\
	\subfloat[Calibration feedback.]{\includegraphics[width =3.1in]{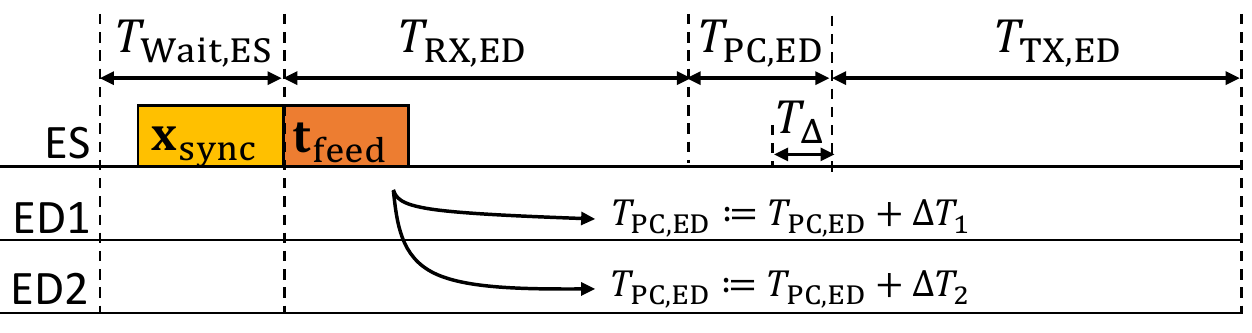}
		\label{subfig:calFeedback}}
	\caption{The proposed procedure for calibration.}
	\label{fig:alignmentProtocol}		
\end{figure}

The baseband processing (and the additional processing for FEEL) at the \ac{ED} can take time in the order of seconds. In this case, $\timePC[]$ may need to be set to a large value accordingly. However, using a large value for $\timePC[]$ (also for $\timeRX[]$ and $\timeTX[]$) results in a surprising time offset problem due to the inaccurate and unstable \ac{FPGA} clock. To elaborate on this, we model the instantaneous \ac{FPGA} clock period $\fpgaClockPeriodActual$ at the $\indexED$th \ac{ED} as	$\fpgaClockPeriodActual = \fpgaClockPeriod + \fpgaClockPeriodOffset + \fpgaClockPeriodNoise$
where $\fpgaClockPeriod$ is the ideal clock period and $\fpgaClockPeriodOffset$ and $\fpgaClockPeriodNoise$ are the offset and the jitter due to the imperfect oscillator on the \ac{SDR}, respectively. The proposed block at the $\indexED$th \ac{ED} measures $\timeRX[,ED]+\timePC[,ED]$ through a counter that counts up till $\Ncount[PC]=(\timeRX[,ED]+\timePC[,ED])/{\fpgaClockPeriod}$ with the \ac{FPGA} clock ticks. Therefore, the difference between $\timeRX[,ED]+\timePC[,ED]$ and the measured one can be calculated as
 \begin{align}
\errorTime[\indexED]=\timePC[]-\sum_{\indexSample=0}^{\Ncount[PC]-1}\fpgaClockPeriodActual={\Ncount[PC] \fpgaClockPeriodOffset} +{\sum_{\indexSample=0}^{\Ncount[PC]-1}  \fpgaClockPeriodNoise}~,
\nonumber
 \end{align} 
which implies that a large $\Ncount[PC]$ causes not only  a large time offset (the first term) but also a large jitter (second term). The jitter can be mitigated by reducing  $\Ncount[PC]$ or using a more stable oscillator in the \ac{SDR}. To address the time offset, we propose a closed-loop calibration procedure as illustrated in \figurename~\ref{fig:alignmentProtocol}. In this method, the \ac{ES} transmits a trigger signal, denoted by ${\cmdPPDU[cal]}$, along with $\ssSync$ as shown in \figurename~\ref{fig:alignmentProtocol}\subref{subfig:calTrigger}. After the $\indexED$th \ac{ED} receives ${\cmdPPDU[cal]}$, it responds to the trigger signal with a calibration signal, denoted by $\signalCal[\indexED]$, $\forall\indexED$, such that the received calibration signals are desired to be aligned back to back. With cross-correlations, the \ac{ES} calculates $\errorTime[\indexED]$, $\forall\indexED$. It then transmits a feedback signal denoted by ${\cmdPPDU[feed]}$ as in \figurename~\ref{fig:alignmentProtocol}\subref{subfig:calFeedback}, where ${\cmdPPDU[feed]}$ contains  time offset information for all \acp{ED}, i.e., $\{\errorTime[\indexED],\forall\indexED\}$. Subsequently, each \ac{ED} updates its local $\timePC[,ED]$ as $\timePC[,ED]+\errorTime[\indexED]$. In this study, we construct ${\cmdPPDU[cal]}$  based on a custom design, detailed in Section~\ref{subsec:signaling}, while the calibration signals are based on \ac{ZC} sequences. 

It is worth noting that the feedback signal may be generalized to include information related received signal power, transmit power increment, or \ac{CFO}. In this study, the feedback signal also contains transmit power offset and \ac{CFO} for each ED so that a coarse power alignment and frequency synchronization can be maintained within the capabilities of the \acp{SDR}.

\section{Proposed OAC Procedure for FEEL}

In this study, we implement \ac{FEEL} based on the \ac{OAC} scheme, i.e., \ac{FSK-MV}, originally proposed in \cite{sahinCommnet_2021} and extended in \cite{sahin2022mv} with the absentee votes. 
To make the reader familiar with this scheme, 
let $\dataset[\indexED]$ denote the local data set containing the labeled data samples $(\sampleData[\indexSampleData], \sampleLabel[\indexSampleData] )$ at the $\indexED$th \ac{ED} for $\indexED=1,\mydots,\numberOfEdgeDevices$, where $\sampleData[\indexSampleData]$ and $\sampleLabel[\indexSampleData]$ are $\indexSampleData$th data sample and its associated label, respectively. 
The main problem tackled with \ac{FEEL} can  be expressed as 
\begin{align}
	\modelParametersOptimal=\arg\min_{\modelParameters} \lossFunctionGlobal[\modelParameters]=\arg\min_{\modelParameters} \frac{1}{|\completeData|}\sum_{\forall(\sampleData[], \sampleLabel[] )\in\completeData} \lossFunctionSample[{\modelParameters,\sampleData[],\sampleLabel[]}]~,
	\label{eq:clp}
\end{align}
where $\completeData=\dataset[1]\cup\cdots\cup\dataset[K]$ and $\lossFunctionSample[{\modelParameters,\sampleData[],\sampleLabel[]}]$ is the sample loss function measuring the labeling error for $(\sampleData[], \sampleLabel[])$ for the parameter vector $\modelParameters=[\modelParametersEle[1],\mydots,\modelParametersEle[\numberOfModelParameters]]^{\rm T}\in\realNumbers^{\numberOfModelParameters}$. 

To solve \eqref{eq:clp} in a wireless network  with \ac{OAC} in a distributed manner (i.e., the global data set $\completeData$ cannot be formed by uploading the local data sets to the \ac{ES}), for the $\indexCommunicationRound$th parameter-update round, the $\indexED$th \ac{ED} first calculates the local stochastic gradients as 
\begin{align}
	\localGradient[\indexED][\indexCommunicationRound] =  \nabla  \lossFunctionLocal[\indexED][{\modelParametersAtIteration[\indexCommunicationRound]}] 
	= \frac{1}{\batchSize} \sum_{\forall(\sampleData[\indexSampleData], \sampleLabel[\indexSampleData] )\in\datasetBatch[\indexED]} \nabla 
	\lossFunctionSample[{\modelParametersAtIteration[\indexCommunicationRound],\sampleData[\indexSampleData],\sampleLabel[\indexSampleData]}]
	~,
	\label{eq:LocalGradientEstimate}
\end{align}
where $\localGradient[\indexED][\indexCommunicationRound]=[\localGradientElement[\indexED,1][\indexCommunicationRound],\mydots,\localGradientElement[\indexED,\numberOfModelParameters][\indexCommunicationRound]]$ is the gradient vector, $\datasetBatch[\indexED]\subset\dataset[\indexED]$ is the selected data batch from the local data set and $\batchSize=|\datasetBatch[\indexED]|$ as the batch size. Similar to the distributed training strategy by the \ac{MV} with \ac{signSGD} \cite{Bernstein_2018}, each \ac{ED} then  activates one of the two subcarriers determined by the time-frequency index pairs $(\voteInTime[+],\voteInFrequency[+])$ and $(\voteInTime[-],\voteInFrequency[-])$  for $\voteInTime[+],\voteInTime[-]\in\{0,1,\mydots,\numberOfOFDMSymbols-1\}$ and $\voteInFrequency[+],\voteInFrequency[-]\in\{0,1,\mydots,\numberOfActiveSubcarriers-1\}$ with the symbols $\transmittedSymbolAtSubcarrier[\indexED,{\voteInFrequency[+]},{\voteInTime[+]}]$ and $\transmittedSymbolAtSubcarrier[\indexED,{\voteInFrequency[-]},{\voteInTime[-]}]$, $\forall\indexGradient$ as
\begin{align}
	\transmittedSymbolAtSubcarrier[\indexED,{\voteInFrequency[+]},{\voteInTime[+]}]=
		\sqrt{\symbolEnergy}\randomSymbolAtSubcarrier[\indexED,\indexGradient]\gradientWeight[{\localGradientElement[\indexED,\indexGradient][\indexCommunicationRound]}]\indicatorFunction[{\sign(\localGradientElement[\indexED,\indexGradient][\indexCommunicationRound]) =1}]
~,
	\label{eq:symbolOne}
\end{align}
and
\begin{align}
	\transmittedSymbolAtSubcarrier[\indexED,{\voteInFrequency[-]},{\voteInTime[-]}]=
	\sqrt{\symbolEnergy}\randomSymbolAtSubcarrier[\indexED,\indexGradient]\gradientWeight[{\localGradientElement[\indexED,\indexGradient][\indexCommunicationRound]}]\indicatorFunction[{\sign(\localGradientElement[\indexED,\indexGradient][\indexCommunicationRound]) =-1}]
	~,
	\label{eq:symbolTwo}
\end{align}
respectively, where  $\gradientWeight[{\localGradientElement[\indexED,\indexGradient][\indexCommunicationRound]}]=1$ for $|\localGradientElement[\indexED,\indexGradient][\indexCommunicationRound]|\ge\thresholdForZero$, otherwise it is $0$, $\symbolEnergy=2$ is the normalization factor,  $\randomSymbolAtSubcarrier[\indexED,\indexGradient]$ is a random \ac{QPSK} symbol to reduce the \ac{PMEPR}, the function $\signNormal[\cdot]$ results in $1$, $-1$, or $\pm1$ at random for a positive, a negative, or a zero-valued argument, respectively, and  the function $\indicatorFunction[\cdot]$ results in $1$ if  its argument holds, otherwise it is $0$. 
The $\numberOfEdgeDevices$ \acp{ED} then access the wireless channel  on the same time-frequency resources {\em simultaneously} with $\numberOfOFDMSymbols$ \ac{OFDM} symbols consisting of $\numberOfActiveSubcarriers$ active subcarriers. In \cite{sahin2022mv}, it is shown that  $\thresholdForZero>0$ (i.e., enabling absentee votes) can improve the test accuracy by eliminating the converging \acp{ED}  from the \ac{MV} calculation when the data distribution is heterogeneous.

Let $	\receivedSymbolAtSubcarrier[{\voteInFrequency[+]},{\voteInTime[+]}]$ and $	\receivedSymbolAtSubcarrier[{\voteInFrequency[-]},{\voteInTime[-]}]$ be  the received symbols after the superposition for the $\indexGradient$th gradient at the \ac{ES}.
The \ac{ES} detects the \ac{MV}  for the $\indexGradient$th gradient with an energy detector as
\begin{align}
	\majorityVoteEle[\indexCommunicationRound][\indexGradient] = \signNormal[{\deltaVectorAtIteration[\indexCommunicationRound][\indexGradient]}]~,
	\label{eq:detector}
\end{align}
where $\deltaVectorAtIteration[\indexCommunicationRound][\indexGradient]\triangleq{\metricForFirst[\indexGradient]-\metricForSecond[\indexGradient]}$  for
$
\metricForFirst[\indexGradient]\triangleq  |\receivedSymbolAtSubcarrier[{\voteInFrequency[+]},{\voteInTime[+]}]|_2^2
$
and
$
\metricForSecond[\indexGradient]\triangleq  |\receivedSymbolAtSubcarrier[{\voteInFrequency[+]},{\voteInTime[+]}]|_2^2$, $\forall\indexGradient$. 
Finally, the \ac{ES} transmits $\majorityVote[\indexCommunicationRound]=[\majorityVoteEle[\indexCommunicationRound][1],\mydots,\majorityVoteEle[\indexCommunicationRound][\numberOfModelParameters]]^{\rm T}$ to the \acp{ED}  and the models at the \acp{ED} are updated as
$	\modelParametersAtIteration[\indexCommunicationRound+1] = \modelParametersAtIteration[\indexCommunicationRound] - \learningRate  \majorityVote[\indexCommunicationRound]$, 
where $\learningRate$ is the learning rate.

In \cite{sahinCommnet_2021} and \cite{sahin2022mv}, the reception of the \ac{MV} vector by the \acp{ED} is assumed to be perfect. In practice, the \acp{MV} can be communicated via traditional communication methods. Nevertheless, as it increases the complexity of the \acp{ED}, we also use the \ac{FSK} in the \ac{DL} in our implementation as done for the \ac{UL}. 

\begin{figure}
	\centering
	\subfloat[Gradient trigger.]{\includegraphics[width =3.1in]{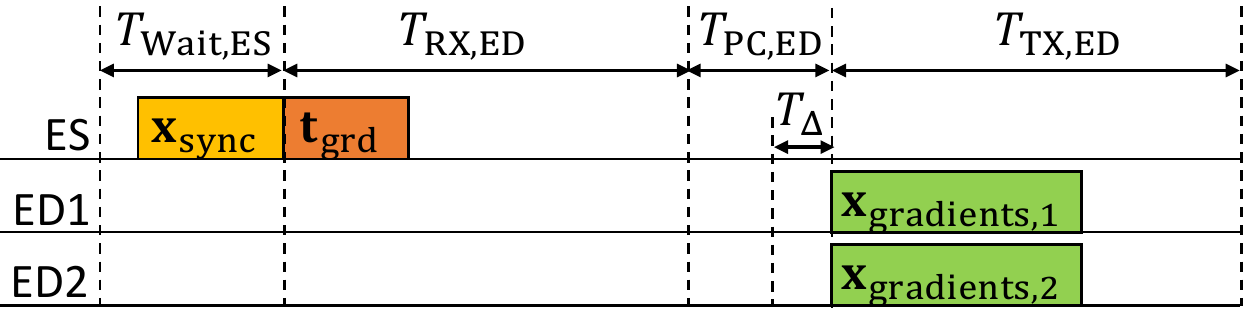}
		\label{subfig:gradientTrigger}}\\
	\subfloat[MV feedback.]{\includegraphics[width =3.1in]{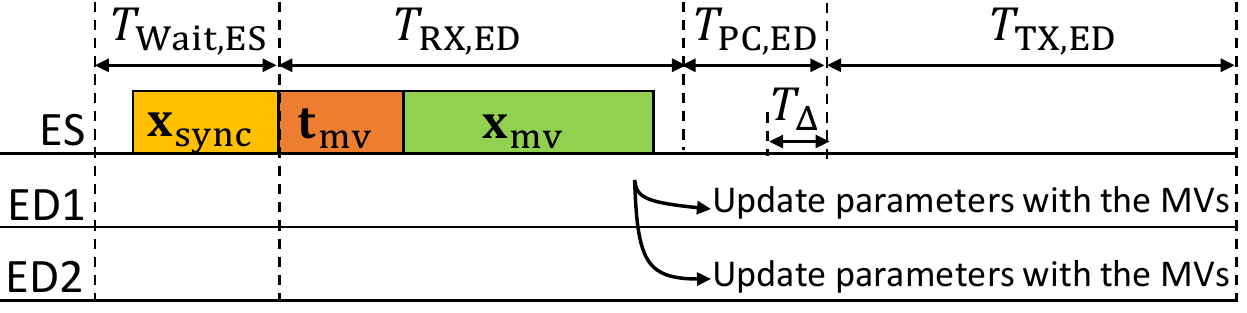}
		\label{subfig:MVfeed}}	
	\caption{The proposed procedure for \ac{OAC} with FSK-MV.}
	\label{fig:OACprotocol}	
\end{figure}
\begin{figure*}[t]
	\centering
	{\includegraphics[width =7in]{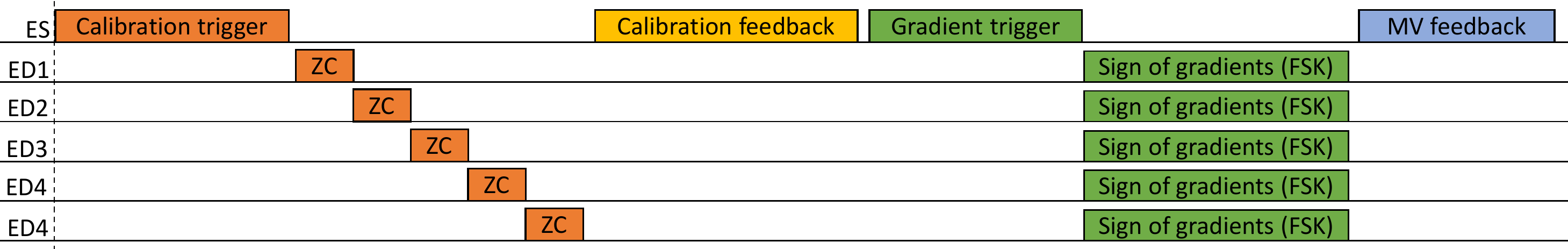}
	} 
	\caption{The procedure for \ac{FSK-MV} along with calibration phase. With the calibration feedback, power offset, \ac{CFO}, and time offset are fed back to each ED.}
	\label{fig:procedure}
\end{figure*}

\def\gradientSignal[#1]{\textbf{x}_{{\rm gradients},#1}}
\def\mvSignal{\textbf{x}_{{\rm mv}}}
In \figurename~\ref{fig:OACprotocol}, we illustrate the proposed procedure for \ac{FSK-MV}. Assuming that the calibration is done via the  procedure in Section~\ref{subsec:nonIdealClocks}, the \ac{ES} initiates the \ac{OAC} by transmitting a trigger signal, i.e., ${\cmdPPDU[grd]}$, along with the synchronization waveform. The $\indexED$th \acp{ED} responds to the received ${\cmdPPDU[grd]}$ with $\gradientSignal[\indexED]$, $\forall\indexED$, i.e., the \ac{IQ} samples calculated based on  \eqref{eq:symbolOne} and \eqref{eq:symbolTwo}. After the ES receives the superposed modulation symbols, it calculates the \acp{MV} with \eqref{eq:detector}, $\forall\indexGradient$. Afterwards, it synthesizes the IQ samples consisting the \ac{OFDM} symbols based on \ac{FSK}, i.e.,  $\mvSignal$ and transmits $\mvSignal$ along with  ${\cmdPPDU[mv]}$ as shown in \figurename~\ref{fig:OACprotocol}\subref{subfig:MVfeed}. Each \ac{ED} decodes ${\cmdPPDU[mv]}$ to detect the following samples include the \acp{MV}. After decoding the received $\mvSignal$, each \ac{ED} updates its model parameters. Similar to ${\cmdPPDU[cal]}$ and ${\cmdPPDU[feed]}$, the signals ${\cmdPPDU[grd]}$ and ${\cmdPPDU[mv]}$ are based on the \ac{PPDU} discussed in Section~\ref{subsec:signaling}. Over all procedure including the calibration phase within one communication round is illustrated in \figurename~\ref{fig:procedure}.

\section{Proposed PPDU for Signaling}
\label{subsec:signaling}

\begin{figure}[t]
	\centering
	{\includegraphics[width =3.5in]{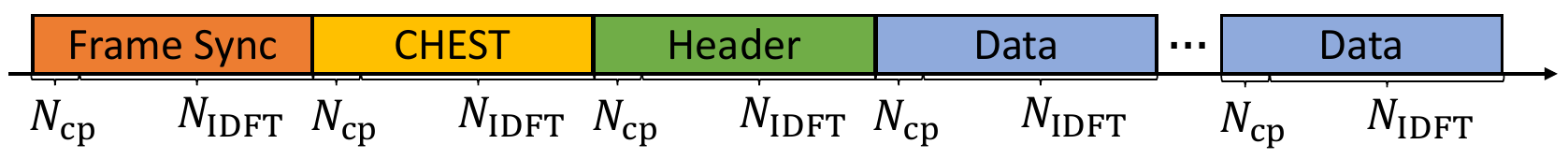}
	} 
	\caption{The structure of the proposed PPDU for  ${\cmdPPDU[cal]}$, ${\cmdPPDU[feed]}$, ${\cmdPPDU[grd]}$, ${\cmdPPDU[mv]}$}
	\label{fig:cmdPPDU}
\end{figure}

\def\golaySequenceA{\textbf{g}_{\rm a}}
\def\golaySequenceB{\textbf{g}_{\rm b}}
\def\numberOfBitsToTransmit{N_{\rm bit}}
\def\numberOfCodewords{N_{\rm cw}}
\def\numberOfPadBits{N_{\rm pad}}
The signaling between \acp{ED} and \ac{ES} in this study is maintained over a custom \ac{PPDU} as shown in \figurename~\ref{fig:cmdPPDU}, and the signaling occurs through the bits transmitted over the \ac{PPDU}. In this design, there are four different fields, i.e., frame synchronization, \ac{CHEST}, header, and data fields, where each field is based on \ac{OFDM} symbols. We express an \ac{OFDM} symbol as
\begin{align}
	\transmittedVector[\indexED,\indexOFDMSymbol] = \cpAddMatrix\idftMatrix[\idftSize]\frequencyMapping \symbolVector[\indexED,\indexOFDMSymbol]~,
	\label{eq:transmitSymbol}
\end{align}
where $\cpAddMatrix\in\realNumbers^{\idftSize+\cpSize\times\idftSize}$ is the \ac{CP} addition matrix, $\idftMatrix[\idftSize]\in\complexNumbers^{\idftSize\times\idftSize}$ is the normalized $\idftSize$-point \ac{IDFT} matrix (i.e., $\idftMatrix[\idftSize]\dftMatrix[\idftSize]=\identityMatrix[\idftSize]$),   $\frequencyMapping\in\realNumbers^{\idftSize\times\numberOfActiveSubcarriers}$ is the mapping matrix that maps the modulation symbols to the subcarriers,  and $\symbolVector[\indexED,\indexOFDMSymbol]\in\complexNumbers^{\numberOfActiveSubcarriers\times1}$ contains the modulation symbols on $\numberOfActiveSubcarriers$ subcarriers. For all fields,
we set the \ac{IDFT} size and the \ac{CP} size as $\idftSize=256$ and $\cpSize=64$, respectively. For \ac{CHEST}, header, and data fields, we use  $\numberOfActiveSubcarriers=192$ active subcarriers along with 8 \ac{DC} subcarriers.  For the frame synchronization field, the \ac{DC} subcarriers are also utilized. 

\subsection{Frame synchronization field}
The frame synchronization field is a single \ac{OFDM} symbol. Every other active subcarrier within the band is utilized with a \ac{ZC} sequence of length $97$. Therefore, the corresponding \ac{OFDM} symbols has two repetitions in the time domain. While the repetitions are used to estimate the \ac{CFO} at the receiver, the null subcarriers are utilized to estimate the noise variance.
\subsection{CHEST field}
The CHEST field is a single \ac{OFDM} symbol. The modulation symbols are the elements of a pair of \ac{QPSK} Golay sequences of length $96$, denoted by $(\golaySequenceA,\golaySequenceB)$. The vector $\symbolVector[]$ is constructed by concatenating $\golaySequenceA$  and $\golaySequenceB$.
\subsection{Header field}
The header is a single \ac{OFDM} symbol. It is based on \ac{BPSK} symbols with a polar code of length $128$ with the rate of $1/2$. We reserve 56 bits for a sequence of signature bits, the number of codewords in the data field, i.e., $\numberOfCodewords$, and the number of pre-padding bits, i.e., $\numberOfPadBits$. The rest of the $8$~bits are reserved for  \ac{CRC}.  We also use \ac{QPSK}-based phase tracking symbols for every other two subcarriers, where the tracking symbols are the elements of a \ac{QPSK} Golay sequence of length $64$.

\subsection{Data field}
Let $\numberOfBitsToTransmit$ be the number of information bits to be communicated. We calculate the number of codewords and the number of pre-padding bits as $\numberOfCodewords=\ceil{\numberOfBitsToTransmit/56}$ and $\numberOfPadBits=56\numberOfCodewords-\numberOfBitsToTransmit$. After the information bits are padded with $\numberOfPadBits$, they are grouped into $\numberOfCodewords$ messages of length 56 bits. The concentration of each message sequence and its corresponding \ac{CRC} is encoded with a polar code  of length $128$ with the rate of $1/2$.  We carry one codeword on each \ac{OFDM} symbol with \ac{BPSK} modulation. Hence, the number of \ac{OFDM} symbols in the data field is also $\numberOfCodewords$. Similar to the header, \ac{QPSK}-based phase tracking symbols are used for every other two subcarriers.

\subsection{Signaling}
Throughout this study, we use the information bits that are transmitted over the \ac{PPDU} to signal ${\cmdPPDU[cal]}$, ${\cmdPPDU[feed]}$, ${\cmdPPDU[grd]}$, ${\cmdPPDU[mv]}$ and user multiplexing. We dedicate $4$ bits for signaling type and $25$~bits for user multiplexing. If the signaling type is the calibration feedback, we define 32 bits for time offset and 8 bits for power control for each \ac{ED}.

\section{Experiment}
\label{sec:numericalResults}

\begin{figure}
	\centering
	\subfloat[The ES with an NVIDIA Jetson Nano.]{\includegraphics[width =0.99in]{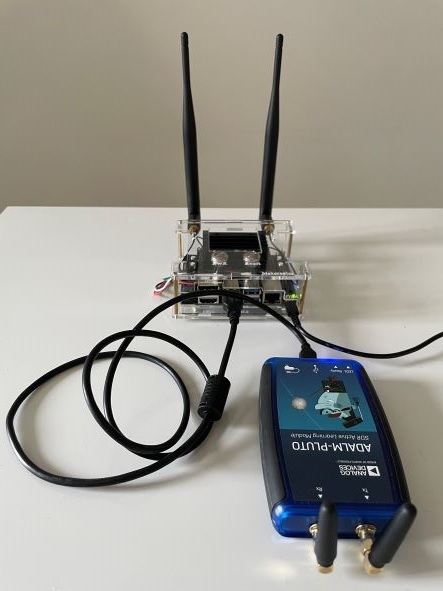}
		\label{subfig:ed}}~
	\subfloat[The \acp{ED} with Microsoft Surface Pro 4. An independent thread runs for each SDR.]{\includegraphics[width =2.4in]{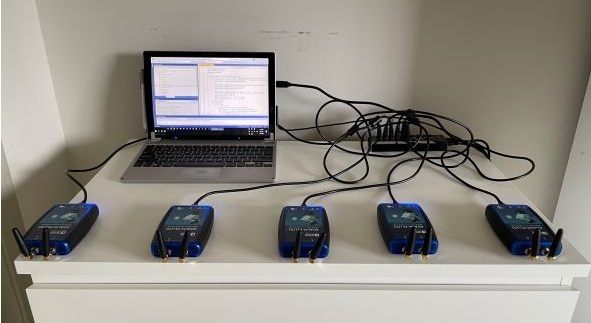}
		\label{subfig:es}}
	\caption{The implemented EDs and ES with an NVIDIA Jetson Nano, a Microsoft Surface Pro 4, and 6 Analog Device Adalm Pluto SDRs for FEEL.}
	\label{fig:experiment}	
\end{figure}

For the experiment, we consider the learning task of handwritten-digit recognition with $\numberOfEdgeDevices=5$ \acp{ED} and \ac{ES}, where each of them is implemented with Adalm Pluto (Rev. C) SDRs.  We develop the \ac{IP} core for the proposed synchronization method by using MATLAB HDL Coder and 
embed it to the FPGA (Xilinx Zynq XC7Z010) based on the guidelines provided in \cite{analogDevicePluto}. As shown in \figurename~\ref{fig:experiment}, we use a Microsoft Surface Pro 4 for the \acp{ED}, where an independent thread runs for each \ac{ED}. The \ac{CC} for the \ac{ES} is an NVIDIA Jetson Nano development module. The baseband and machine learning algorithms are written in the Python language. We run the experiment in an indoor environment where the mobility is relatively low. The link distance between an \ac{ED} and the \ac{ES} is approximately 5~m. The sample rate is $\sampleRate=20$~Msps for all radios and the signal bandwidth is approximately $15$~MHz.   We synthesize the vectors ${\cmdPPDU[cal]}$, ${\cmdPPDU[feed]}$, ${\cmdPPDU[grd]}$, ${\cmdPPDU[mv]}$ based on the custom \ac{PPDU} discussed in Section~\ref{subsec:signaling} and consider the same \ac{OFDM} symbol structure in the PPDU for $\mvSignal$  $\gradientSignal[\indexED]$, and $\signalCal[\indexED]$, $\forall\indexED$. For the synchronization IP, the pre-configured values of $\timeWait[,ES]$, $\timePC[,ED]$, $\timeRX[,ED]$, $\timeTX[,ED]$, and $\timeDelta$ are $750$~ms, $750$~ms, $50$~ms, $50$~ms, and $100~\mu$s, respectively.

We use the MNIST database that contains labeled handwritten-digit images. To prepare the data, we first choose $|\completeData|=25000$ training images from the database, where each digit has distinct $2500$ images.  
For homogeneous data distribution, each \ac{ED} has $500$ distinct images for each digit. For heterogeneous data distribution, $\indexED$th \ac{ED} has the data samples with the labels $\{\indexED-1,\indexED,1+\indexED,2+\indexED,3+\indexED,4+\indexED\}$. For both distributions, the EDs do not have common training images.  For the model, we consider a \ac{CNN} that consists of two 2D convolutional layers with the kernel size $[5, 5]$, stride  $[1, 1]$, and padding $[2, 2]$, where the former layer has $1$ input and $16$ output channels and the latter one has $16$ input and $32$ output channels. Each layer is followed by batch norm, rectified linear units, and max pooling layer with the kernel size 2. Finally, we use a fully-connected layer followed by softmax. Our model has $\numberOfModelParameters=29034$ learnable parameters that result in $\numberOfOFDMSymbols=303$ OFDM symbols for \ac{FSK-MV}. We set $\learningRate=0.001$ and $\batchSize=100$. For the test accuracy, we use $10000$ test samples in the database.

\begin{figure}[t]
	\centering
	{\includegraphics[width =3.5in]{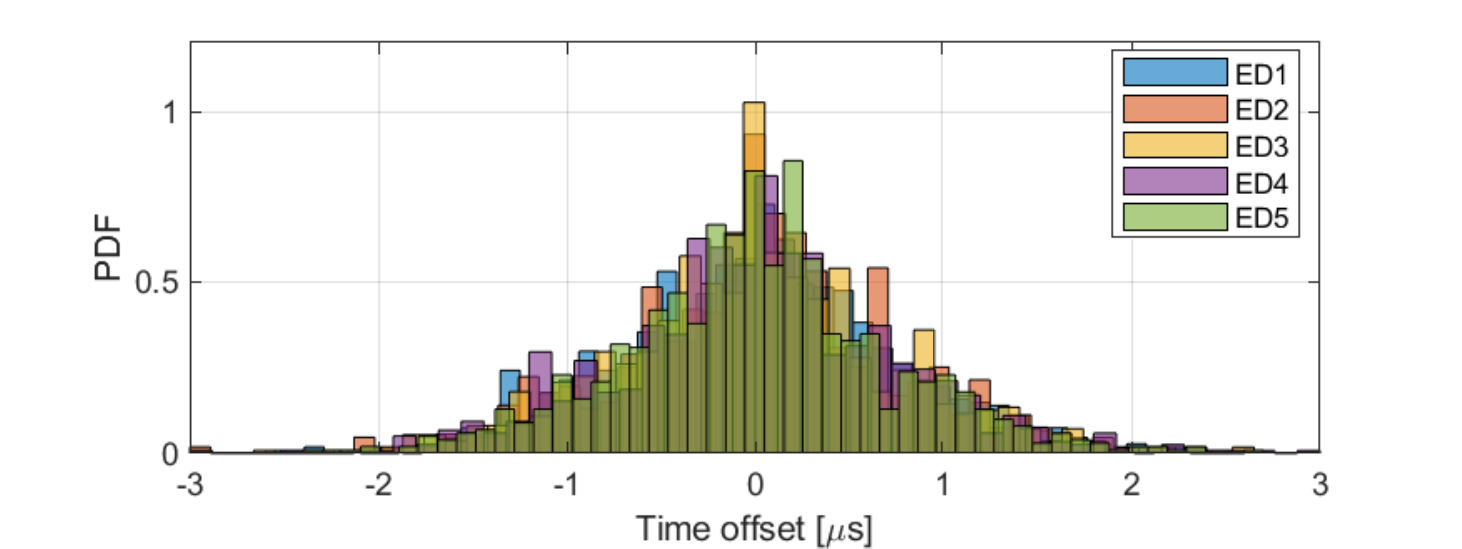}
	} 
	\caption{The distribution of time synchronization error  due to the imperfect clocks in Adalm Pluto SDRs.}
	\label{fig:toDist}
\end{figure}
\begin{figure}[t]
	\centering
	{\includegraphics[width =3.3in]{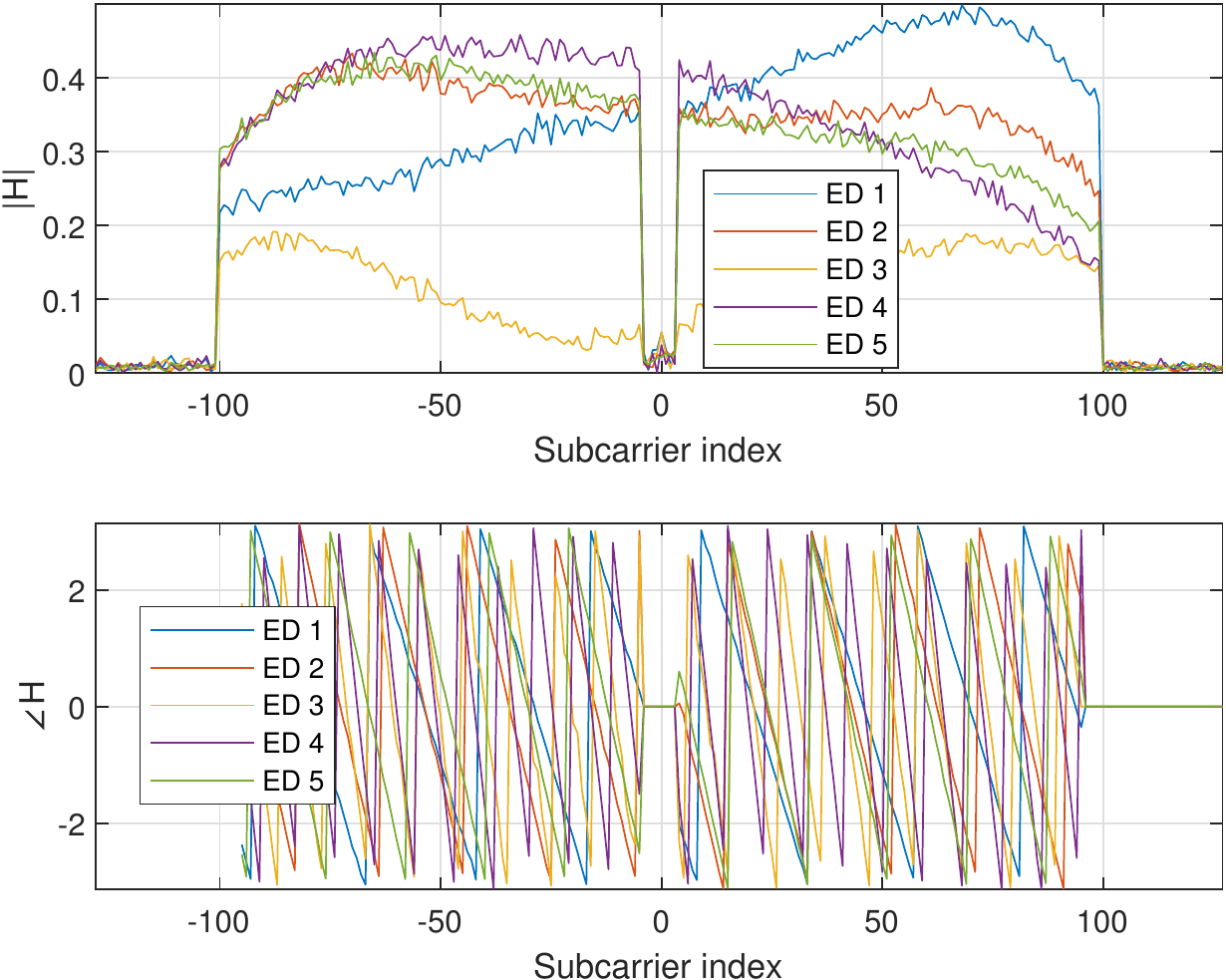}
	} 
	\caption{An instant of channel frequency response during the experiment. }
	\label{fig:channel}
\end{figure}
The experiment reveals many practical issues. The \ac{FPGA} clock rate of Adalm Pluto SDR is 100~MHz, generated from a 40 MHz oscillator where the frequency deviation is rated at 20 PPM. Due to the large deviation and $\timePC[,ED]+\timeRX[,ED]$, we observe a large time offset and a large jitter as discussed in Section~\ref{subsec:nonIdealClocks}. Hence, the \ac{ES} initiates the calibration procedure in \figurename~\ref{fig:alignmentProtocol} after completing the \ac{OAC} procedure in \figurename~\ref{fig:OACprotocol}, sequentially.
In \figurename~\ref{fig:toDist}, we provide the distribution of the jitter after the calibration, where the standard deviation of the jitter is around $1~\mu$s for $\timePC[,ED]+\timeRX[,ED]=0.8$~s. Although the jitter can be considerably large, we conduct the experiment under this impairment to demonstrate the robustness of \ac{FSK-MV} against synchronization errors. In the experiment, a line-of-sight path is present. Nevertheless, the channel between an \ac{ED} and the \ac{ES} is still frequency selective as can be seen in \figurename~\ref{fig:channel}. In the experiment, we observe that the magnitudes of the channel frequency coefficients do not change significantly due to the low mobility. However, their phases change in an intractable manner due to the random time offsets. Nevertheless, this is not an issue for \ac{FSK-MV} as it does not require a phase alignment. Note that we also implement a closed-loop power control by using the calibration procedure to align the received signal powers. However, an ideal power alignment is challenging to maintain. For example, ED 3's channel in \figurename~\ref{fig:channel} is relatively under a deep fade, but the \ac{SDR}'s transmit power cannot be increased further. Similar to the jitter, we run the experiment under non-ideal power control.

\def\figuresize{1.725in}
\begin{figure}
	\centering
	\subfloat[Homegenous, wo. absentee votes.]{\includegraphics[width =\figuresize]{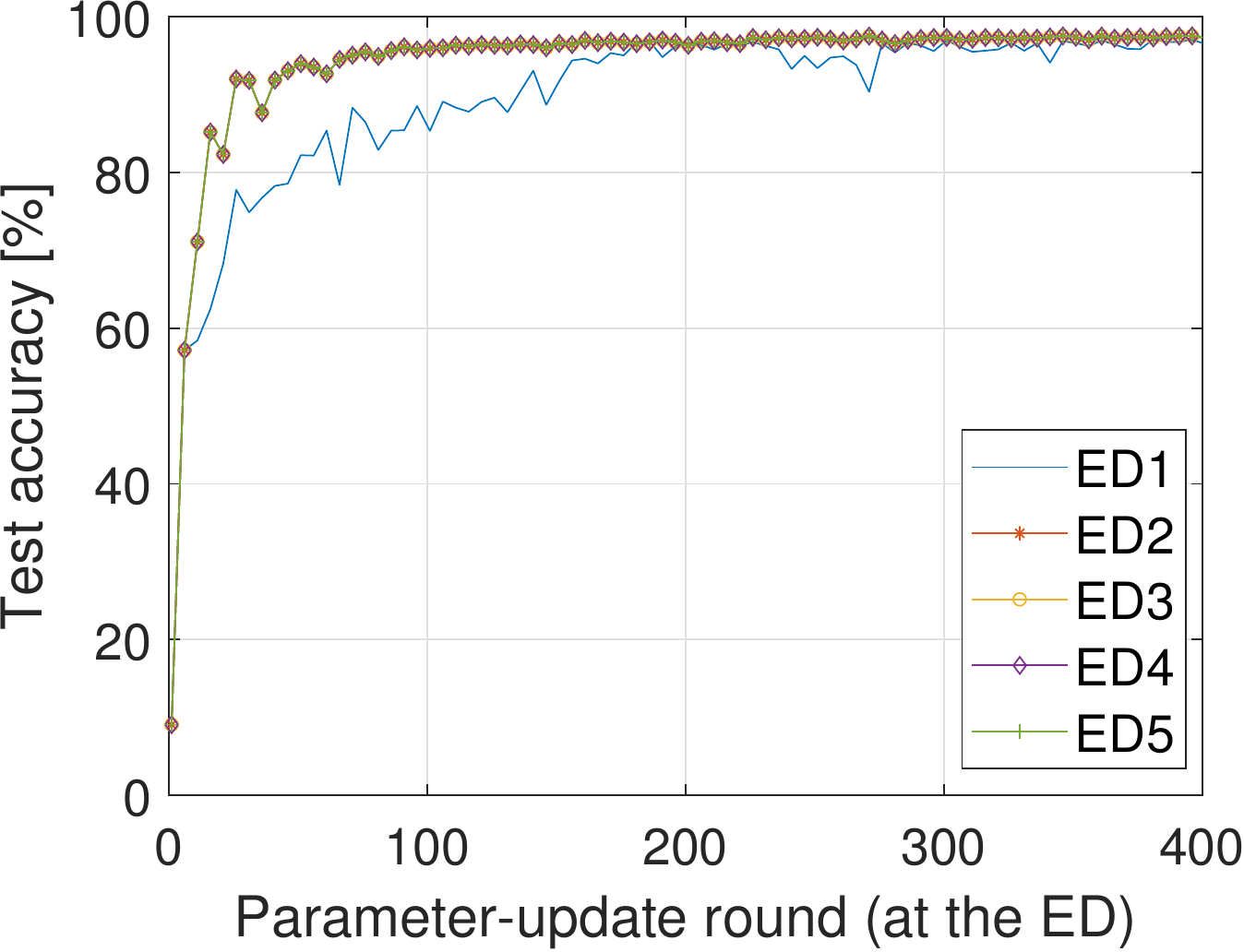}
		\label{subfig:acc_hete0_abs0}}	
	\subfloat[Homegenous, w. absentee votes.]{\includegraphics[width =\figuresize]{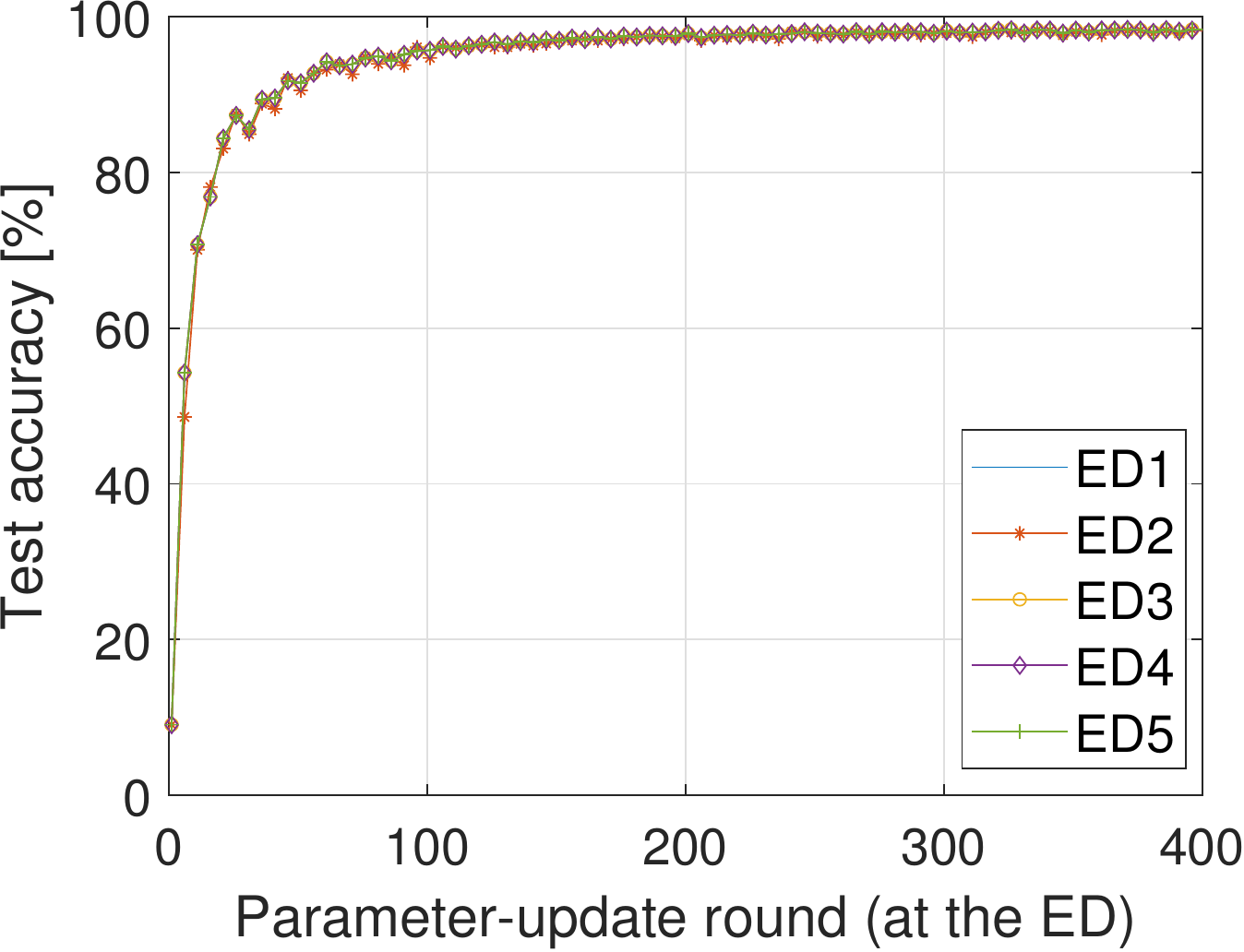}
		\label{subfig:acc_hete0_abs1}}	
	\\
	\subfloat[Homegenous, wo. absentee votes.]{\includegraphics[width =\figuresize]{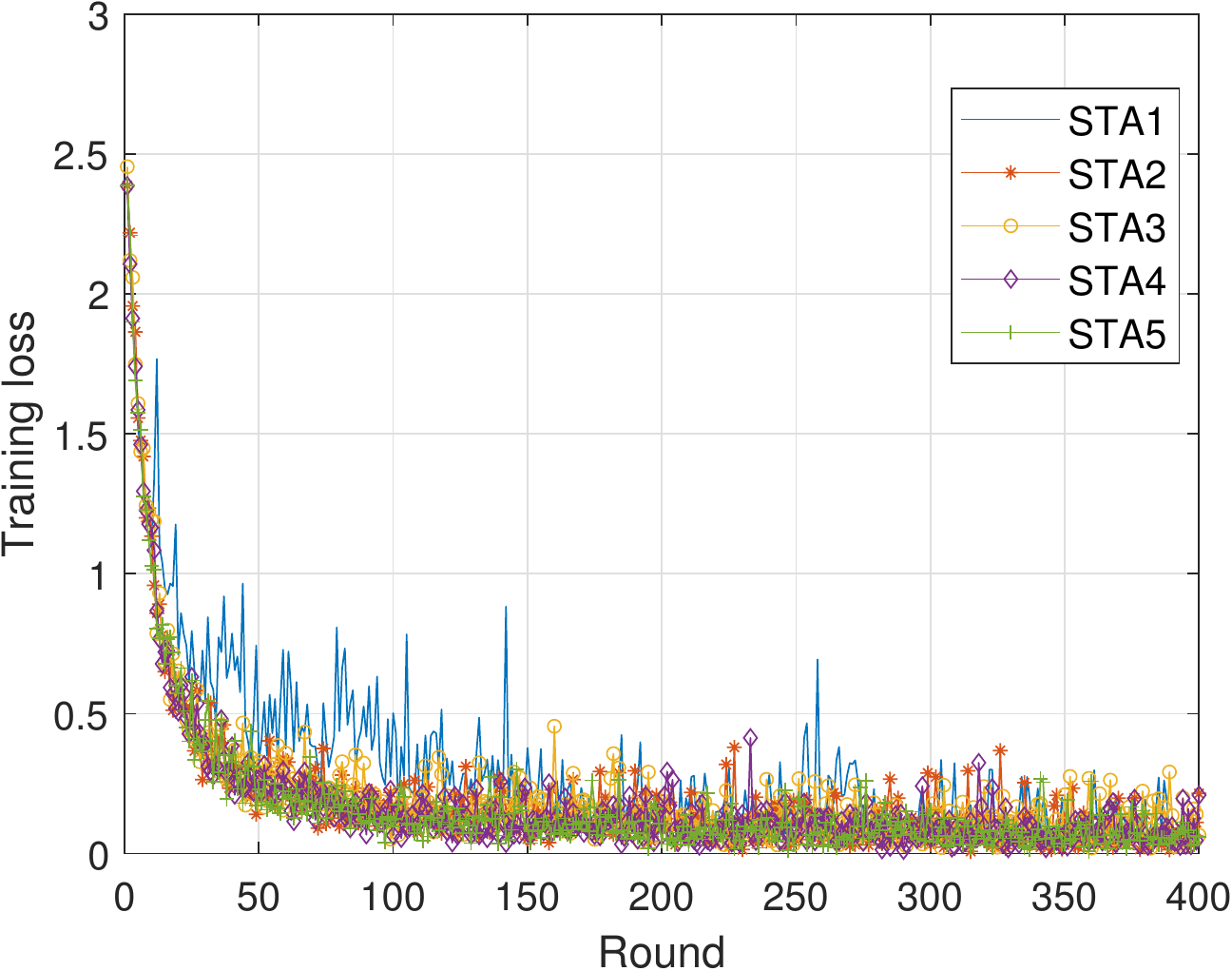}
		\label{subfig:loss_hete0_abs0}}			
	\subfloat[Homegenous, w. absentee votes.]{\includegraphics[width =\figuresize]{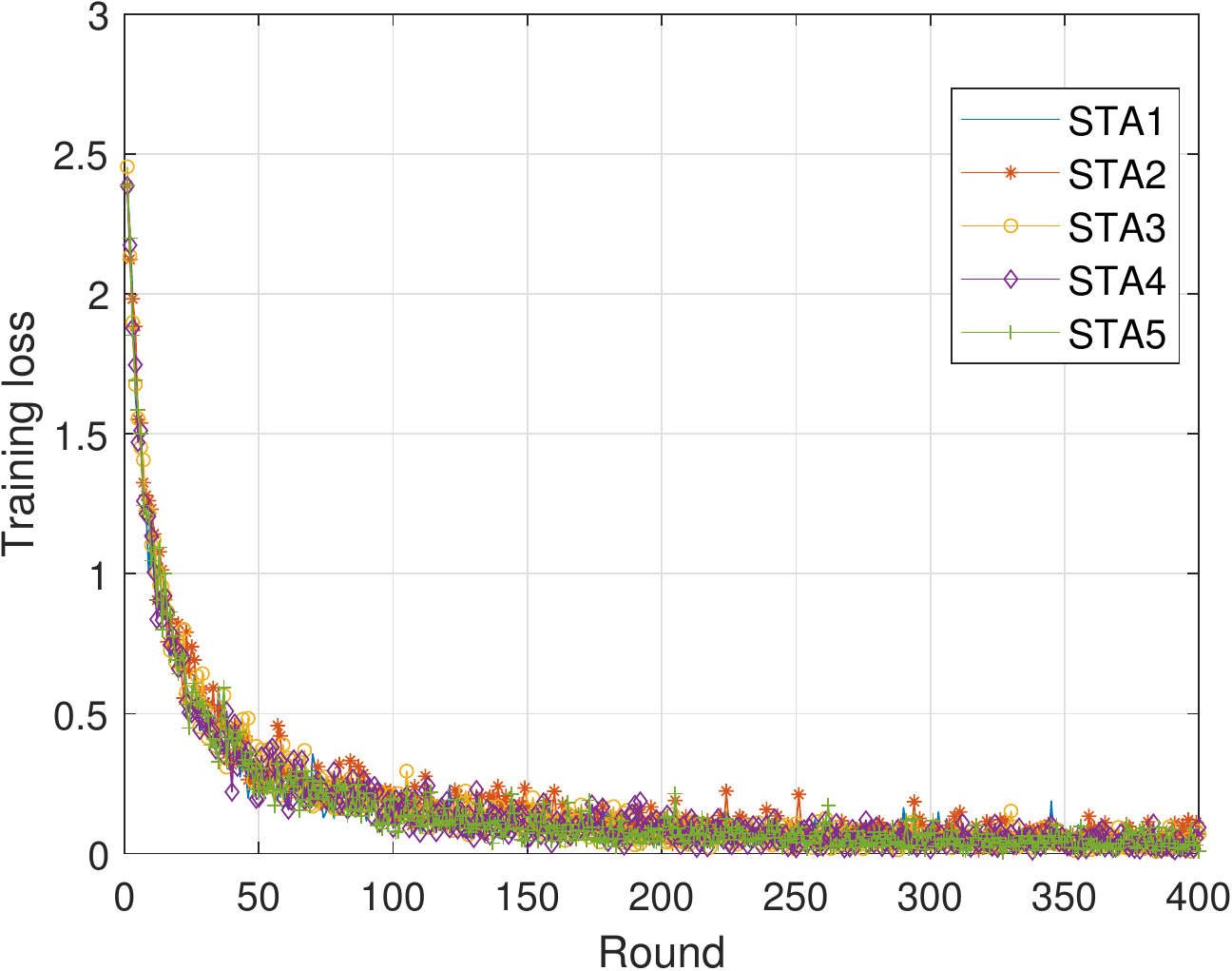}
		\label{subfig:loss_hete0_abs1}}	
	\\	
	\subfloat[Heterogeneous, wo. absentee votes.]{\includegraphics[width =\figuresize]{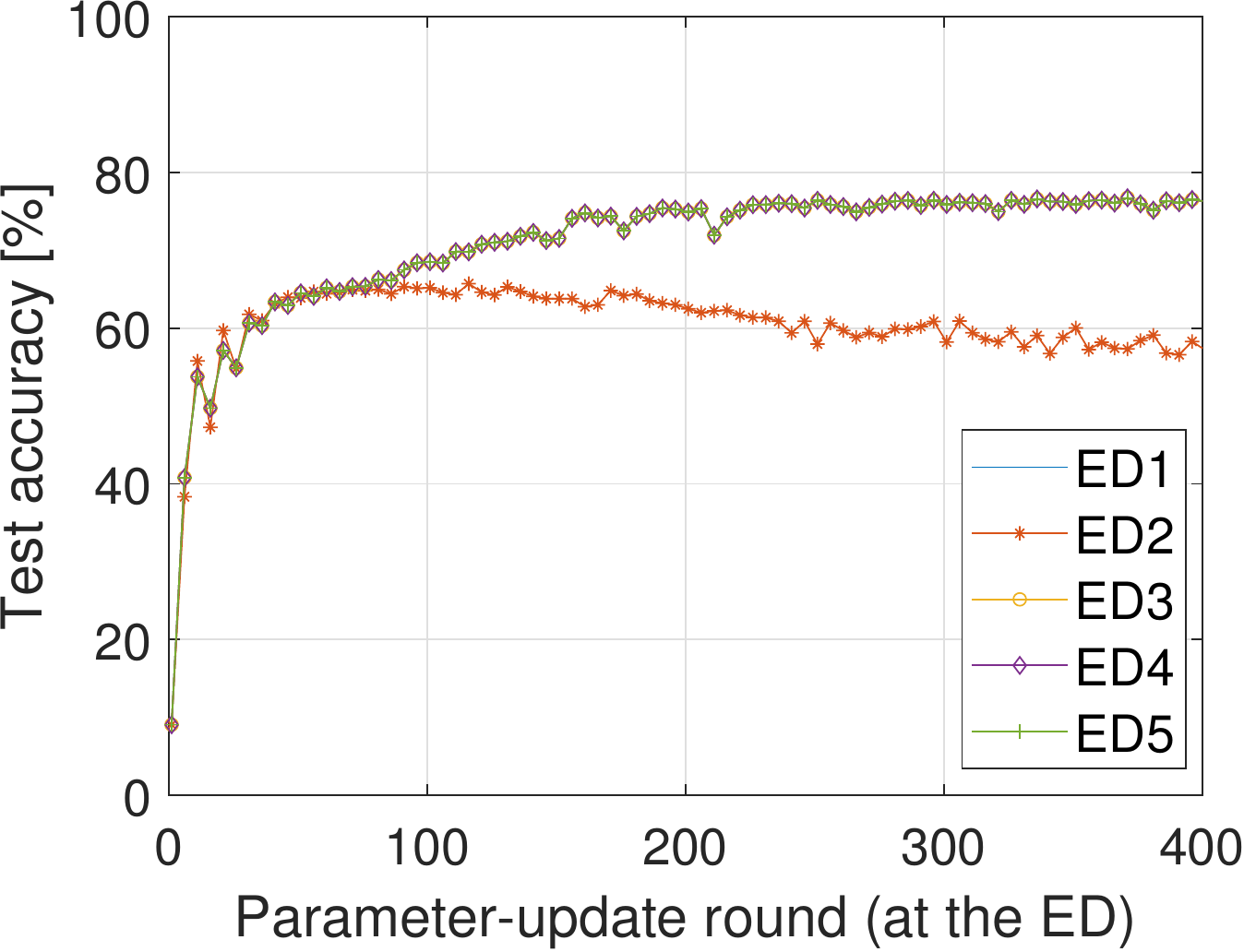}
		\label{subfig:acc_hete1_abs0}}	
	\subfloat[Heterogeneous, w. absentee votes.]{\includegraphics[width =\figuresize]{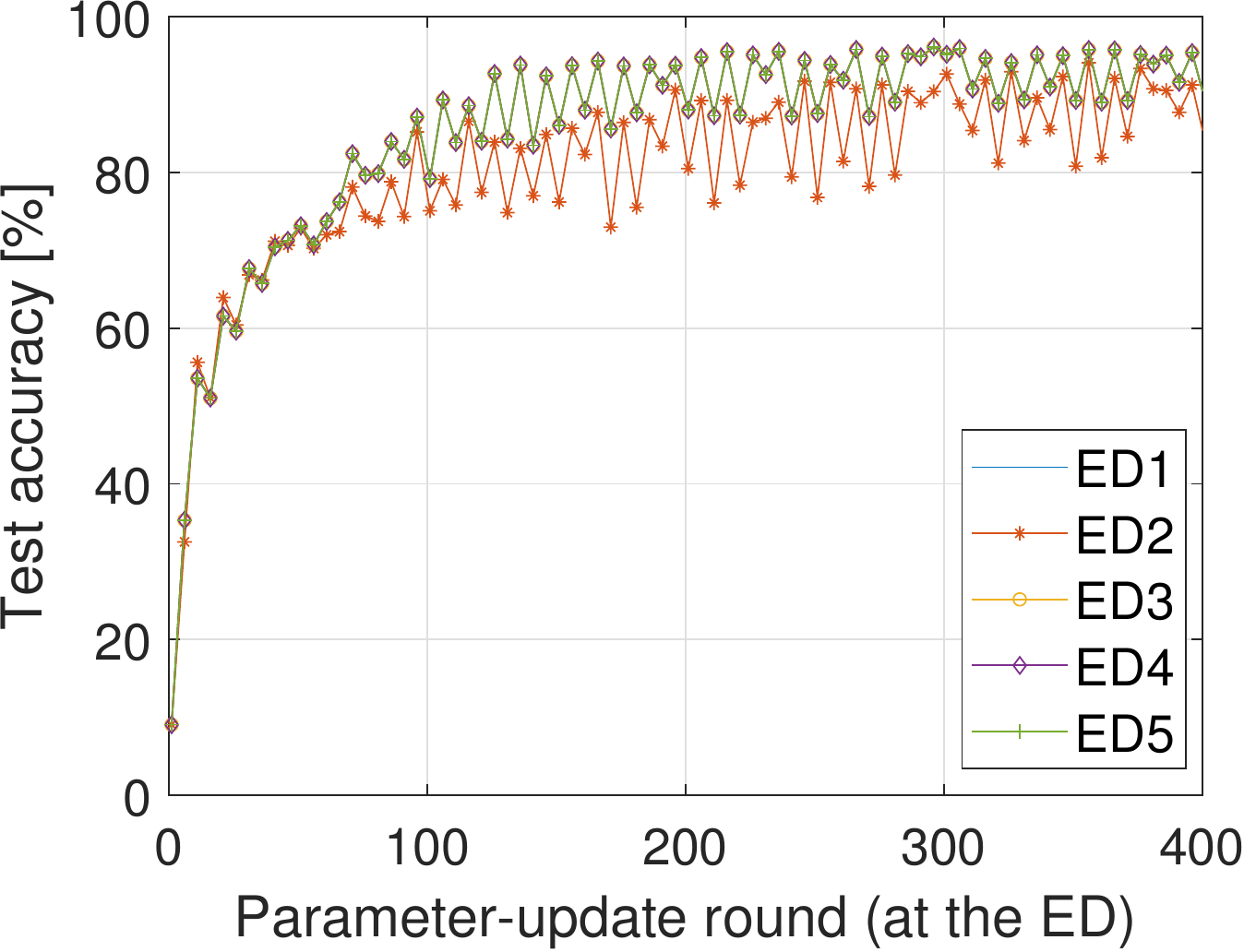}
		\label{subfig:acc_hete1_abs1}}			
	\\
	\subfloat[Heterogeneous, wo. absentee votes.]{\includegraphics[width =\figuresize]{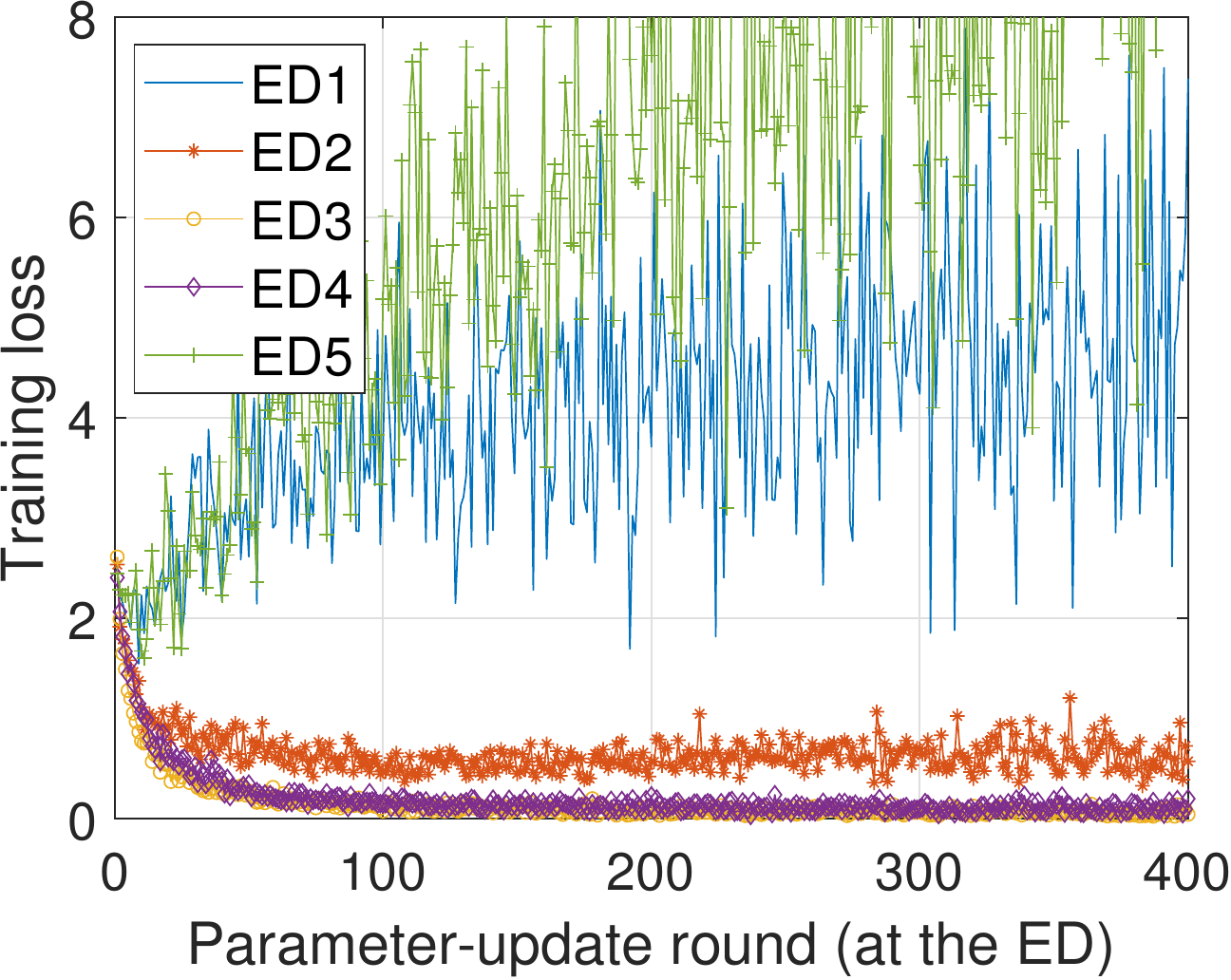}
		\label{subfig:loss_hete1_abs0}}			
	\subfloat[Heterogeneous, w. absentee votes.]{\includegraphics[width =\figuresize]{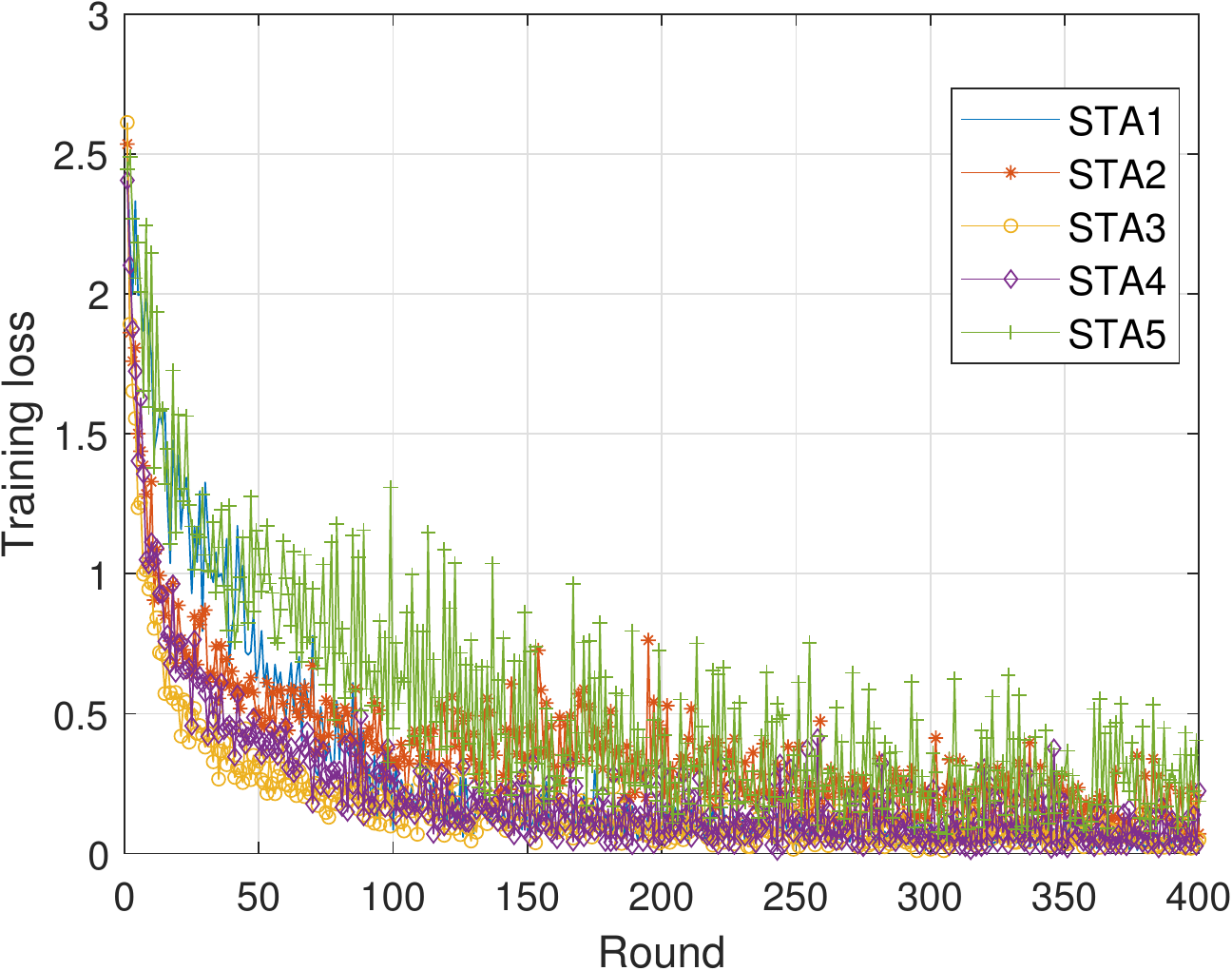}
		\label{subfig:loss_hete1_abs1}}		
	\caption{Experiment results for the FEEL with the OAC scheme FSK-MV with/without absentee votes.}
	\label{fig:experimentResults}
\end{figure}
Finally, in \figurename~\ref{fig:experimentResults}, we provide the test accuracy and training loss at each \ac{ED} when the training is done without  absentee votes ($\thresholdForZero=0$) and with absentee votes ($\thresholdForZero=0.005$). For homogeneous data distribution, the test accuracy for each ED quickly reaches 97.5\% for both cases as given in \figurename~\ref{fig:experimentResults}\subref{subfig:acc_hete0_abs0} and \figurename~\ref{fig:experimentResults}\subref{subfig:acc_hete0_abs1}. The corresponding training losses also decrease as shown in \figurename~\ref{fig:experimentResults}\subref{subfig:loss_hete0_abs0} and \figurename~\ref{fig:experimentResults}\subref{subfig:loss_hete0_abs1}. For heterogeneous data distribution scenario, eliminating converging \ac{ED} improves the test accuracy considerably. For example,  in \figurename~\ref{fig:experimentResults}\subref{subfig:loss_hete1_abs0}, the training losses for ED~1 and ED~5  gradually increase, which indicates that the digit $0$ and the digit $9$ cannot be learned well since these images are available at ED~1 and ED~5. Therefore, the test accuracy drops below 80\% as shown in \figurename~\ref{fig:experimentResults}\subref{subfig:acc_hete1_abs0}. However, with absentee votes, this issue is largely addressed and test accuracy reaches 95\% as can be seen in \figurename~\ref{fig:experimentResults}\subref{subfig:acc_hete1_abs1}.

\section{Conclusion}
In this study, we propose a method that can maintain the synchronization in an \ac{SDR}-based network without implementing the baseband as a hard-coded block. We also provide the corresponding procedure and discuss the design of the synchronization waveform to address the hardware limitations. Finally, by implementing the proposed concept with Adalm Pluto SDRs, for the first time, we demonstrate the performance of an \ac{OAC}, i.e., \ac{FSK-MV}, for \ac{FEEL}. Our experiment shows that \ac{FSK-MV} provides robustness against time synchronization errors and can result in a high test accuracy in practice.

\acresetall

\bibliographystyle{IEEEtran}
\bibliography{references}

\end{document}